\documentclass[11pt]{article}
\usepackage{fullpage}
\usepackage{amsfonts, amsthm, amssymb, amsmath}
\usepackage[pdftex]{graphicx,color}
\usepackage[scaled=0.92]{helvet}
\usepackage{times}

\usepackage[pdftex, pagebackref=false, colorlinks=true, urlcolor=black, citecolor=black, linkcolor=black]{hyperref}

\usepackage{framed}

\usepackage{fullpage}

\usepackage{epsfig}

\usepackage{verbatim}

\usepackage{algorithm}
\usepackage{algorithmic}

\newtheorem{thm}{Theorem}[section]
\newtheorem{lem}[thm]{Lemma}
\newtheorem{cor}[thm]{Corollary}
\newtheorem{defn}[thm]{Definition}
\newtheorem{clm}[thm]{Claim}
\newtheorem{cons}[thm]{Construction}

\newenvironment{theorem}{\begin{thm}\begin{rm}}%
{\end{rm}\end{thm}}
\newenvironment{lemma}{\begin{lem}\begin{rm}}%
{\end{rm}\end{lem}}
\newenvironment{corollary}{\begin{cor}\begin{rm}}%
{\end{rm}\end{cor}}
\newenvironment{definition}{\begin{defn}\begin{em}}%
{\end{em}\end{defn}}
{\end{rm}\end{clm}}
{\end{em}\end{cons}}

\newcommand{\secref}[1]{\hyperref[#1]{Section \ref{#1}}}
\newcommand{\apref}[1]{\hyperref[#1]{Appendix \ref{#1}}}
\newcommand{\thref}[1]{\hyperref[#1]{Theorem \ref{#1}}}
\newcommand{\defref}[1]{\hyperref[#1]{Definition \ref{#1}}}
\newcommand{\corref}[1]{\hyperref[#1]{Corollary \ref{#1}}}
\newcommand{\lemref}[1]{\hyperref[#1]{Lemma \ref{#1}}}
\newcommand{\clref}[1]{\hyperref[#1]{Claim \ref{#1}}}
\newcommand{\consref}[1]{\hyperref[#1]{Construction \ref{#1}}}
\newcommand{\figref}[1]{\hyperref[#1]{Figure \ref{#1}}}
\newcommand{\eqnref}[1]{\hyperref[#1]{Equation \ref{#1}}}

\newcommand{\LOGDCFL}{${\bf{LogDCFL}}$}

\newcommand{\coLOGCFL}{${\bf{co}}$-${\bf{LogCFL}}$}
\newcommand{\coSLOGCFL}{${\bf{co}}$-${\bf{SLogCFL}}$}

\newcommand{\LOGCFL}{${\bf{LogCFL}}$}
\newcommand{\SLOGCFL}{${\bf{SLogCFL}}$}
\newcommand{\SGSLOGCFL}{${\bf{SGSLogCFL}}$}

\newcommand{\oneLOGCFL}{${\bf{1LogCFL}}$}
\newcommand{\oneSLOGCFL}{${\bf{1SLogCFL}}$}
\newcommand{\oneSGSLOGCFL}{${\bf{1SGSLogCFL}}$}

\newcommand{\BPLOGCFL}{${\bf{BPLogCFL}}$}
\newcommand{\RLOGCFL}{${\bf{RLogCFL}}$}

\newcommand{\ZPLOGCFL}{${\bf{ZPLogCFL}}$}

\newcommand{\logspace}{${\bf{L}}$}
\newcommand{\NL}{${\bf{NL}}$}
\newcommand{\CONL}{${\bf{co}}$-${\bf{NL}}$}
\newcommand{\RL}{${\bf{RL}}$}
\newcommand{\BPL}{${\bf{BPL}}$}
\newcommand{\SL}{${\bf{SL}}$}

\newcommand{\NCone}{${\bf{{NC}^{1}}}$}
\newcommand{\NCtwo}{${\bf{{NC}^{2}}}$}
\newcommand{\SCtwo}{${\bf{{SC}^{2}}}$}

\newcommand{\SACone}{${\bf{{SAC}^{1}}}$}

\newcommand{\parityNL}{${\bf{{\oplus}NL}}$}
\newcommand{\parityLOGCFL}{${\bf{{\oplus}LogCFL}}$}

\newcommand{\STCONN}{{\sc{ST-Connectivity}}}
\newcommand{\USTCONN}{{\sc{Undirected ST-Connectivity}}}

\newcommand{\BSTCONN}{{\sc{Balanced ST-Connectivity}}}
\newcommand{\PBSTCONN}{{\sc{Positive Balanced ST-Connectivity}}}

\newcommand{\kBSTCONN}{{\sc{k-Balanced ST-Connectivity}}}
\newcommand{\kPBSTCONN}{{\sc{Positive k-Balanced ST-Connectivity}}}

\newcommand{\STREAL}{{\sc{ST-Realizability}}}
\newcommand{\USTREAL}{{\sc{Undirected ST-Realizability}}}
\newcommand{\SGUSTREAL}{{\sc{Symmetric Gap Undirected ST-Realizability}}}

\newcommand{\randreduce}{$\leq_r$}

\title{{\bf{Realizable Paths and the \NL\ vs \logspace\ Problem}}}
\vspace{0.15in}
\author{
{Shiva Kintali} \\
\vspace{0.10in} \\
College of Computing, \\
Georgia Institute of Technology, \\
Atlanta, GA 30332-0765. \\
{\small \href{mailto:kintali@cc.gatech.edu}{\nolinkurl{kintali@cc.gatech.edu}}}
}
\date{}

\begin{document}

\maketitle

\begin{abstract}
A celebrated theorem of Savitch \cite{Savitch70} states that $NSPACE(S)$ $\subseteq$ $DSPACE(S^2)$. In particular, Savitch gave a deterministic algorithm to solve \STCONN\ (an \NL-complete problem) using $O({\log}^2{n})$ space, implying \NL\ $\subseteq$ $DSPACE({\log}^2{n})$. While Savitch's theorem itself has not been improved in the last four decades, studying the space complexity of several special cases of \STCONN\ has provided new insights into the space-bounded complexity classes.

In this paper, we introduce new kind of graph connectivity problems which we call {\em{graph realizability problems}}. All of our graph realizability problems are generalizations of \USTCONN. \STREAL, the most general graph realizability problem, is \LOGCFL-complete. We define the corresponding complexity classes that lie between \logspace\ and \LOGCFL\ and study their relationships.

As special cases of our graph realizability problems we define two natural problems, \BSTCONN\ and \PBSTCONN, that lie between \logspace\ and \NL. We present a deterministic $O({\log}{n}{\log}{\log}{n})$ space algorithm for \BSTCONN. More generally we prove that \SGSLOGCFL, a generalization of \BSTCONN, is contained in $DSPACE({\log}{n}{\log}{\log}{n})$. To achieve this goal we generalize several concepts (such as graph squaring and transitive closure) and algorithms (such as parallel algorithms) known in the context of \USTCONN. \\

\end{abstract}

\thispagestyle{empty}
\newpage
\setcounter{page}{1}

\section{Introduction}

A celebrated theorem of Savitch \cite{Savitch70} states that $NSPACE(S)$ $\subseteq$ $DSPACE(S^2)$. In particular, Savitch gave a deterministic algorithm to solve \STCONN\ (an \NL-complete problem) using $O({\log}^2{n})$ space, implying \NL\ $\subseteq$ $DSPACE({\log}^2{n})$. Savitch's algorithm runs in time $2^{O({\log}^2{n})}$. It has been a longstanding open problem to improve Savitch's theorem i.e., to prove (i) \NL\ $\subseteq {DSPACE}(o({\log}^{2}{n}))$ or (ii) \NL\ $\subseteq$ \SCtwo, i.e., \STCONN\ can be solved by a deterministic algorithm in polynomial time and $O({\log}^2{n})$ space.

While Savitch's theorem itself has not been improved in the last four decades, studying the space complexity of several special cases of \STCONN\ has provided new insights into the space-bounded complexity classes. Allender's survey \cite{allender-stconn-survey} gives an update of progress related to several special cases of \STCONN. Recently \STCONN\ in planar DAGs with $O({\log}{n})$ sources is shown to be in \logspace\ \cite{planar-few-sources}. Stolee and Vinodchandran proved that \STCONN\ in DAGs with $2^{O(\sqrt{{\log}n})}$ sources embedded on surfaces of genus $2^{O(\sqrt{{\log}n})}$ is in \logspace\ \cite{reach-surface-embedded}.

All the connectivity problems considered in the literature so far are essentially special cases of \STCONN. In the first half of this paper, we introduce new kind of graph connectivity problems which we call {\em{graph realizability problems}}. All of our graph realizability problems are generalizations of \USTCONN. \STREAL, the most general graph realizability problem is \LOGCFL-complete. We define the corresponding complexity classes that lie between \logspace\ and \LOGCFL\ and study their relationships. As special cases of our graph realizability problems we define two natural problems, \BSTCONN\ and \PBSTCONN, that lie between \logspace\ and \NL.

In the second half of this paper, we study the space complexity of \SGSLOGCFL\ (see \secref{subsec:sgslogcfl} for definition). We define generalizations of graph squaring and transitive closure, present efficient parallel algorithms for \SGSLOGCFL\ and use the techniques of Trifonov \cite{trifonov-logloglog} to show that \SGSLOGCFL\ is contained in $DSPACE({\log}{n}{\log}{\log}{n})$. This implies that \BSTCONN, a natural graph connectivity problem which lies between \logspace\ and \NL, is contained in $DSPACE({\log}{n}{\log}{\log}{n})$.

\subsection{Preliminaries, Related Work and Our Results}

\noindent {\bf{Auxiliary Pushdown Automata}} : A language is accepted by a non-deterministic pushdown automaton (PDA) if and only if it is a context-free language. Deterministic context-free languages are those accepted by the deterministic PDAs. \LOGCFL\ is the set of all languages that are log-space reducible to a context-free language. Similarly, \LOGDCFL\ is the set of all languages that are log-space reducible to a deterministic context-free language. There are many equivalent characterizations of \LOGCFL. Sudborough \cite{Sudborough78} gave the machine class equivalence. Ruzzo \cite{Ruzzo80} gave an alternating Turing machine (ATM) class equivalent to \LOGCFL. Venkateswaran \cite{Venkateswaran91} gave a circuit characterization and showed that \LOGCFL\ = \SACone. For a survey of parallel complexity classes and \LOGCFL\ see Limaye's thesis \cite{nutan-logcfl}.

An Auxiliary Pushdown Automaton (NAuxPDA or simply AuxPDA), introduced by Cook \cite{cook-auxpda}, is a two-way PDA augmented with an $S(n)$-space bounded work tape. If a deterministic two-way PDA is augmented with an $S(n)$-space bounded work tape then we get a Deterministic Auxiliary Pushdown Automaton (DAuxPDA). We present the formal definitions in the {\bf{appendix}} (see Section \ref{sec:symmauxpda}). Let {\em{NAuxPDA-SpaceTime}} ($S(n)$,$T(n)$) be the class of languages accepted by an AuxPDA with $S(n)$-space bounded work tapes and the running time bounded by $T(n)$. Let the corresponding deterministic class be {\em{DAuxPDA-SpaceTime}} ($S(n)$,$T(n)$). It is easy to see that \NL $\ \subseteq\ $ {\em{NAuxPDA-SpaceTime}} ($O({\log}n)$, $poly(n)$). It is shown by Sudborough that {\em{NAuxPDA-SpaceTime}} ($O({\log}n)$, $poly(n)$) = \LOGCFL\ and {\em{DAuxPDA-SpaceTime}} ($O({\log}n)$,$poly(n)$) = \LOGDCFL\ \cite{Sudborough78}. Using ATM simulations, Ruzzo showed that \LOGCFL\ $\subseteq$ \NCtwo\ \cite{Ruzzo80}. Simpler proofs of {\em{DAuxPDA-SpaceTime}} ($O({\log}n)$,$poly(n)$) = \LOGDCFL\ and \LOGCFL\ = \SACone\ are given in \cite{circuits-cfls}.

Many proof techniques and results obtained in the context of \NL, are generalized to obtain the corresponding results for \LOGCFL. For example : (i) Borodin \cite{borodin-nl-nc2} proved that \NL\ $\subseteq$ \NCtwo. Ruzzo \cite{Ruzzo80} introduced tree-size-bounded alternating Turing machines, gave a new characterization of \LOGCFL, and proved that \LOGCFL\ $\subseteq$ \NCtwo. (ii) Immerman \cite{Immerman88} and Szelepcs\'{e}nyi \cite{Szelepcsenyi87} proved that \NL\ = \CONL. Borodin et. al. \cite{BCDRT89} generalized their inductive counting technique and proved that \LOGCFL\ = \coLOGCFL. In fact, they proved a stronger result showing that ${\bf{SAC}}^i$ is closed under complementation for $i>0$. (iii) Wigderson \cite{Wigderson-parityNL} proved that \NL\ \randreduce\ \parityNL. G{\'a}l and Wigderson \cite{GalWigderson96} proved that \LOGCFL\ \randreduce\ \parityLOGCFL. (iv) Nisan \cite{nisan-rlsc} proved that \BPL\ $\subseteq$ \SCtwo. Venkateswaran \cite{venkat-auxpda, venkat-auxpda-techreport} proved that \BPLOGCFL\ $\subseteq$ \SCtwo\ and \BPLOGCFL\ $\subseteq$ \NCtwo. Here \BPLOGCFL\ (resp. \RLOGCFL\ and \ZPLOGCFL) is the bounded error (resp. one-sided error and zero error) probabilistic version of \LOGCFL. All the above results are elegant and non-trivial generalizations of the corresponding results in the logspace setting.

Throughout this paper, we consider $O({\log}n)$-space bounded and polynomial-time bounded AuxPDAs. The {\em{surface configuration}} (introduced by Cook \cite{cook-auxpda}) of an AuxPDA, on an input $w$, consists of the state, contents and head positions of the work tapes, the head position of the input tape and the topmost symbol of the stack i.e., the rightmost symbol of the pushdown tape. Note that for an $S(n)$-space bounded AuxPDA, its surface configurations take only $O(S(n))$ space. In the rest of the paper, we will refer to surface configurations as configurations. For an input $w$, a pair of configurations $(C_1,C_2)$ is {\em{realizable}} if the AuxPDA can move from $C_1$ to $C_2$ ending with its stack at the same height as in $C_1$, and without popping its stack below its level in $C_2$ for any of the intermediate configurations. An AuxPDA $M$ accepts an input $w$ iff there is a realizable pair $(I,A)$, where $I$ is the initial configuration and $A$ is the unique accepting configuration. \\

\noindent {\bf{Realizable Paths}} : \STCONN\ (resp. \USTCONN) is the problem of determining whether there exists a path between two distinguished vertices $s$ and $t$ in a directed (resp. undirected) graph. These two graph connectivity problems played a central role in understanding the complexity classes \logspace, \SL\ and \NL\ \cite{AKLLR, SymmLogspace, BCDRT89, ustconn-log3by2, KWspan, SL=coSL, sakszhou3by2, ustconn-log4by3, zigzag, trifonov-logloglog, SL=L}.

In \secref{sec:paths}, we introduce a new graph connectivity problem, which we call \STREAL\ and prove that \STREAL\ is complete for \LOGCFL.  \STREAL\ is a generalization of \STCONN, which is \NL-complete. Our definition of \STREAL\ is motivated by (i) Hardest CFL \cite{hardestCFL, Sudborough78, harrison-book}, (ii) Labeled Acyclic GAP, which is \LOGCFL-complete \cite{greenlaw-book} (iii) CFL-reachability, which is ${\bf{P}}$-complete \cite{CFL-reachability, CFL-Pcomplete-PODS, CFL-Pcomplete-reps, CFL-Pcomplete-FOCS} and (iv) the insights from Niedermeier and Rossmanith's parsimonious simulation of \LOGCFL\ by \SACone\ circuits \cite{NiedermeierR95}.

Unlike \STCONN, using breadth-first search (or) depth-first search and keeping track of ``visited" vertices does not result in a polynomial time algorithm for \STREAL. In \secref{sec:transitive}, we generalize the notions of transitive closure and graph squaring. Using these generalizations we present a natural polynomial time algorithm to compute the generalized transitive closure, thus solving \STREAL. \\

\noindent {\bf{Symmetric AuxPDAs}} : In \secref{sec:ustconn}, we define \USTREAL, a ``symmetric" version of \STREAL. To study the space complexity of \USTREAL\ we define {\em{symmetric}} auxiliary pushdown automata, a natural generalization of symmetric Turing machines introduced by Lewis and Papadimitriou \cite{SymmLogspace}. We introduce a new complexity class called \SLOGCFL, a generalization of \SL\ and show that \LOGDCFL\ $\subseteq$ \SLOGCFL\ $\subseteq$ \LOGCFL. \\

\noindent {\bf{Graph Realizability Problems}} : In \secref{sec:realproblems}, we study several variants of \STREAL\ and the corresponding complexity classes. All of these complexity classes lie between \logspace\ and \LOGCFL. In particular, \BSTCONN\ and \PBSTCONN\ are natural graph connectivity problems that lie between \logspace\ and \NL. Figure \ref{fig:real-classes} summarizes the relationship among the newly defined classes. \\

\noindent {\bf{Space Efficient Algorithms}} : The \logspace\ vs \SL\ question (i.e., is there a log space algorithm for solving \USTCONN) motivated an exciting series of new concepts and techniques. Prior to the work of Lewis and Papadimitriou \cite{SymmLogspace}, Aleliunas et. al. \cite{AKLLR} proved that \USTCONN\ $\in$ \RL, implying \SL\ $\subseteq$ \RL. Nisan, Szemeredi and Wigderson \cite{ustconn-log3by2} showed that \USTCONN\ can be solved deterministically in space $O({\log}^{\frac{3}{2}}{n})$. This result was later subsumed by a beautiful result of Saks and Zhou, showing that ${BP}_{H}{SPACE}({S}) \subseteq {DSPACE}({S}^{3/2})$ \cite{sakszhou3by2}. Armoni, et. al. \cite{ustconn-log4by3} showed that \USTCONN\ $\in DSPACE({\log}^{\frac{4}{3}}{n})$. Trifonov \cite{trifonov-logloglog} gave an $O({\log}{n}{\log}{\log}{n})$-space deterministic algorithm for \USTCONN. Independently at the same time, using completely different techniques, Reingold \cite{SL=L} settled the space complexity of \USTCONN\ and proved that \SL\ = \logspace. The zig-zag graph product, introduced by Reingold, Vadhan and Wigderson \cite{zigzag-journal}, played a crucial role in Reingold's algorithm.

Our space efficient algorithm for \SGSLOGCFL\ (see \secref{sec:sgslogcfl-logloglog}) is based on Trifinov's technique \cite{trifonov-logloglog}, which is based on Chong-Lam's parallel algorithm \cite{parallel-ustconn-lognloglogn} solving \USTCONN\ in $O({\log}{n}{\log}{\log}{n})$ time on EREW PRAM. This necessitates the development of such a parallel algorithm for \SGSLOGCFL. \\

\noindent {\bf{Parallel Algorithms}} : Hirschberg, Chandra and Sarwate \cite{parallel-ustconn-log2n} presented an $O({\log}^2{n})$ time parallel algorithm using $n^2/{\log}{n}$ processors on a CREW PRAM to find connected components of an undirected graph. Their algorithm remained the best known for almost a decade. In a breakthrough work, Johnson and Metaxas \cite{parallel-ustconn-log3by2n} presented a CREW algorithm running in $O({\log}^{\frac{3}{2}}{n})$ time using $n+m$ processors. Subsequently they improved their algorithm to run on an EREW PRAM with the same time complexity and number of processors \cite{parallel-ustconn-log3by2n-erew}. Chong and Lam \cite{parallel-ustconn-lognloglogn} presented an $O({\log}{n}{\log}{\log}{n})$ time deterministic EREW PRAM algorithm with $O(m + n)$ processors. Chong, Han, and Lam \cite{parallel-ustconn-logn} showed that the problem can be solved on the EREW PRAM in $O({\log}{n})$ time with $O(m + n)$ processors.

In \secref{sec:parallel-sgslogcfl}, we generalize the algorithms of \cite{parallel-ustconn-log2n}, \cite{parallel-ustconn-log3by2n} and \cite{parallel-ustconn-lognloglogn} and design the corresponding parallel algorithms for \SGSLOGCFL. In \secref{sec:sgslogcfl-logloglog}, we use these algorithms to prove that \SGSLOGCFL\ is contained in $DSPACE({\log}{n}{\log}{\log}{n})$.

\section{Realizable Paths}\label{sec:paths}

\subsection{\STREAL}

We are given a directed graph $\mathcal{G}(V,E)$, a vertex labeling function $L_{V}:V{\rightarrow}\{\alpha_1,\alpha_2,\dots,\alpha_k\}$ and an edge labeling function $L_{E}:E{\rightarrow}\{push,pop,\epsilon\}$. The ordered pair $(s,t)$, where $s,t \in V$, is said to be {\bf{realizable}} if the following two conditions hold :
\begin{itemize}
\item{There is a directed path (say $P$) from $s$ to $t$.}
\item{The concatenation of the vertex and edge labels along the path $P$ is a {\em{realizable}} string (see \defref{defn:realstring}).}
\end{itemize}

\begin{definition}\label{defn:realstring}
Let $\mathcal{A} = \{push,pop,\epsilon,\alpha_1,\alpha_2,\dots,\alpha_k\}$ be the set of alphabets. A {\bf{realizable string}} is a nonempty string of alphabets from $\mathcal{A}$, defined in the following recursive manner :
\begin{itemize}
\item{for all $1 \leq i \leq k$, ``$\alpha_i$" is a realizable string.}
\item{for all $1 \leq i \leq k$, ``${\alpha_i}\ \epsilon\ {\alpha_i}$" is a realizable string.}
\item{if $S$ is a realizable string then so is ``${\alpha_i}\ push\ S\ pop\ {\alpha_i}$", for all $1 \leq i \leq k$.}
\item{for all $1 \leq i \leq k$, if ``${\alpha_i}\ S_1\ {\alpha_i}$" and ``${\alpha_i}\ S_2\ {\alpha_i}$" are realizable strings then so is ``${\alpha_i}\ S_1\ {\alpha_i}\ S_2\ {\alpha_i}$".}
\end{itemize}
\end{definition}

\begin{framed}
\noindent \STREAL\ : Given a directed graph $\mathcal{G}(V,E)$ with vertices labeled from $\{\alpha_1,\alpha_2,\dots,\alpha_k\}$ and edges labeled from $\{push,pop,\epsilon\}$ and two distinguished nodes $s$ and $t$, decide if there is a realizable path from $s$ to $t$ in $\mathcal{G}$.
\end{framed}

We use the notation $(u{\leadsto}v)$ to denote that there is a realizable path from $u$ to $v$. If all the vertices of $\mathcal{G}$ are labeled $\alpha_1$ (i.e., $k=1$) and all the edges are labeled $\epsilon$, we get an instance of \STCONN. Hence, \STREAL\ is a generalization of \STCONN.

\begin{theorem}\label{thm:streal}
\STREAL\ is \LOGCFL-complete.
\end{theorem}

\begin{corollary}\label{cor:strealnoepsilon}
\STREAL\ with no $\epsilon$-edges is \LOGCFL-complete.
\end{corollary}

\subsection{Graph Representation}

We now discuss the representation of an instance of \STREAL\ i.e., a directed graph $\mathcal{G}$ with the vertex and edge labels. Let this graph be $\mathcal{G}(V,E)$ with $|V|=n$. For simplicity we assume that there are no multi-edges. We represent $\mathcal{G}$ as a 4-tuple $\mathcal{G} = {\langle}\mathcal{L},\mathcal{P}_{push},\mathcal{P}_{pop},\mathcal{E}{\rangle}$, where $\mathcal{L}$ is an integer array of length $n$, $\mathcal{P}_{push}$, $\mathcal{P}_{pop}$ and $\mathcal{E}$ are $n{\times}n$ boolean matrices. $\mathcal{L}$ is an integer array of length $N$ representing the vertex labels. $\mathcal{L}[u]$ represents the label of vertex $u$ i.e., $\mathcal{L}[u] = i$ iff the label of $u$ is $\alpha_i$. The $[u,v]^{th}$ entry of the matrix $\mathcal{P}_{push}$ (resp. $\mathcal{P}_{pop}$ and $\mathcal{E}$) is 1 if and only if the directed edge $(u,v)$ is labeled {\em{push}} (resp. {\em{pop}} and $\epsilon$). We may assume that $L_E(u,u) = \epsilon$ for all $u \in V$ i.e., $\mathcal{E}[u,u] = \epsilon$ for all $u \in V$.

\subsection{Gap Matrix}\label{subsec:gapmatrixdefn}

\begin{definition}
{\em{(Niedermeier and Rossmanith \cite{NiedermeierR95})}} : Let {\em{a,b,c,d}} be four configurations such that : {\em{a}} and {\em{b}} have same pushdown heights, {\em{c}} and {\em{d}} have same pushdown heights and there exists a computation path from {\em{a}} to {\em{c}} and one from {\em{d}} to {\em{b}}. The level of the pushdown must not go below the level of {\em{a}} and {\em{b}} during the computation. We say that {\em{(a,b)}} is {\em{realizable with gap}} {\em{(c,d)}}.
\end{definition}

In the context of \STREAL, we relax the above definition as shown below. This allows us to define a natural repeated squaring algorithm to solve \STREAL. For the rest of this paper, we will use the following definition.

\begin{framed}
\noindent {\bf{Path with gap}} : A {\em{path with gap}} consists of four vertices $a,b,c,d$ such that (i) there is a computation path $P_1$ from $a$ to $c$ and $P_2$ from $d$ to $b$ (ii) the vertex labels of $a$ and $b$ are the same (iii) the vertex labels of $c$ and $d$ are the same (iv) let $P$ be the path formed by concatenating $P_1$ and $P_2$ i.e., identifying $c$ and $d$ (iv) the concatenation of the vertex and edge labels along the path $P$ is a {\em{realizable}} string. We denote such a ``path with gap" by $(a{\leadsto}(c,d){\leadsto}b)$ and say that {\em{(a,b)}} is {\em{realizable with gap}} {\em{(c,d)}}.
\end{framed}

{\em{Pair-with-gap}} $(a{\leadsto}(c,d){\leadsto}b)$ is interpreted as if the two surface configurations $c$ and $d$ were the same, i.e., as if a realizable path from $c$ to $d$ would exist. To keep track of paths with gaps, we maintain a boolean {\em{gap matrix}} $\Upsilon$, indexed by 4-tuple of vertices $[a,(c,d),b]$ such that if $\Upsilon[a,(c,d),b]= 1$ then $(a{\leadsto}(c,d){\leadsto}b)$. We initialize the gap matrix $\Upsilon$ with the labels from the matrices $\mathcal{L}$,$\mathcal{P}_{push}$ and $\mathcal{P}_{pop}$ as follows.

\noindent \line(1,0){400} \\
\noindent ${\bf{Initialize Gap Matrix}}({\Upsilon})$ \\
\indent for all $a,b,c,d \in V$\ \ \ $\Upsilon[a,(c,d),b]=0$ \\
\indent for all $a,b,c,d \in V$ \\
\indent \indent {\bf{if}} $((\mathcal{P}_{push}[a,c]==1){\&}{\&}(\mathcal{P}_{pop}[d,b]==1){\&}{\&}(\mathcal{L}[a]==\mathcal{L}[b]){\&}{\&}(\mathcal{L}[c]==\mathcal{L}[d]))$ \\
\indent \indent \indent {\bf{then}} ${\Upsilon}[a,(c,d),b] = 1$ \\
\indent for all $a \in V$\ \ \ $\Upsilon[a,(a,a),a]=1$ \\
\indent for all $a,b \in V$\ \ \  $\Upsilon[a,(a,b),b]=1$ \\
\noindent \line(1,0){400} \\

All the required information from the matrices $\mathcal{L}$,$\mathcal{P}_{push}$ and $\mathcal{P}_{pop}$ is now present in the gap matrix ${\Upsilon}$. Note that we are implicitly removing the ``unnecessary" edges as follows. \\

\noindent {\bf{Removing unnecessary edges}} : If $s$ and $t$ are realizable in ${\mathcal{G}}$ along a path $P$ then the {\em{push}} and {\em{pop}} edges along $P$ have to ``match" i.e., every {\em{push}} label has a corresponding {\em{pop}} label. In other words, if there is a {\em{push}} edge $(a,c)$ such that the label of $a$ is $\alpha_i$ and the label of $c$ is $\alpha_j$ then there is a corresponding {\em{pop}} edge $(d,b)$ along the path $P$ such that the label of $d$ is $\alpha_j$ and the label of $b$ is $\alpha_i$. Hence, we can remove the unnecessary edges as follows :
\begin{itemize}
\item{Let $(u,v)$ be a {\em{push}} edge in $\mathcal{G}$ such that the label of $u$ is $\alpha_i$ and the label of $v$ is $\alpha_j$. If there is no {\em{pop}} edge in $\mathcal{G}$ (other than $(v,u)$) with the vertex labels $(\alpha_j,\alpha_i)$, then remove the edge $(u,v)$.}
\item{Let $(u,v)$ be a {\em{pop}} edge in $\mathcal{G}$ such that the label of $u$ is $\alpha_i$ and the label of $v$ is $\alpha_j$. If there is no {\em{push}} edge in $\mathcal{G}$ (other than $(v,u)$) with the vertex labels $(\alpha_j,\alpha_i)$, then remove the edge $(u,v)$.}
\end{itemize}

We call $\mathcal{E}$ the {\em{standard}} matrix and $\Upsilon$ the {\em{gap}} matrix and assume that an instance of \STREAL, $\mathcal{H}$, is represented by an $n \times n$ standard matrix $\mathcal{E}$ and an $n^2 \times n^2$ gap matrix ${\Upsilon}$ and denote this by $\mathcal{H} = {\langle}{\Upsilon},\mathcal{E}{\rangle}$. The rows and columns of ${\Upsilon}$ are indexed by pairs of vertices of $\mathcal{H}$. $\Upsilon[a,(c,d),b]$ corresponds to the $[(a,b),(c,d)]^{th}$ entry in the $n^2 \times n^2$ matrix.

\section{\USTREAL\ and Symmetric AuxPDAs}\label{sec:ustconn}

\subsection{\USTREAL}

We are given an undirected graph $\mathcal{G}(V,E)$, a vertex labeling function $L_{V}:V{\rightarrow}\{\alpha_1,\alpha_2,\dots,\alpha_k\}$ and an edge labeling function $L_{E}:E{\rightarrow}\{push,pop,\epsilon\}$. Moreover, the edge labels are ``symmetric" i.e., they satisfy the following properties : (i) $L_E(u,v) = push$ if and only if $L_E(v,u) = pop$ and (ii) $L_E(u,v) = \epsilon$ if and only if $L_E(v,u) = \epsilon$.

The pair $(s,t)$, where $s,t \in V$, is said to be {\em{realizable}} if there is an undirected path (say $P$) from $s$ to $t$ and the concatenation of the vertex and edge labels along the path $P$ is a {\em{realizable}} string. Since the edge labels are symmetric, $(s,t)$ is realizable if and only if $(t,s)$ is realizable. We denote this by $(s{\leftrightsquigarrow}t)$.

\begin{framed}
\noindent \USTREAL\ : Given an undirected graph $\mathcal{G}(V,E)$ with vertices labeled from $\{\alpha_1,\alpha_2,\dots,\alpha_k\}$ and {\em{symmetric}} edge labels from $\{push,pop,\epsilon\}$ and two distinguished nodes $s$ and $t$, decide if $s$ and $t$ are realizable in $\mathcal{G}$.
\end{framed}

If all the vertices of $\mathcal{G}$ are labeled $\alpha_1$ (i.e., $k=1$) and all the edges are labeled $\epsilon$, we get an instance of \USTCONN. Hence, \USTREAL\ is a generalization of \USTCONN. To study the space complexity of \USTREAL\ we introduce {\em{symmetric}} AuxPDAs in the following subsection.

\subsection{Symmetric AuxPDAs}

Intuitively, a {\em{symmetric}} AuxPDA is a nondeterministic multi-tape Turing machine which has an extra tape called pushdown tape, with an additional requirement that every move of the machine is ``reversible". In other words, the ``yields" relation between its (surface) configurations is symmetric. Such a machine is allowed to scan two symbols at a time on each of its tapes. We present the formal definitions, properties of symmetric AuxPDAs and the proofs of the following theorems in the {\bf{appendix}} (see \apref{sec:symmauxpda}). We define \SLOGCFL\ to be the class of languages accepted by a log space bounded and polynomial time bounded symmetric AuxPDA.

\begin{theorem}\label{thm:gen-SL-inclusion}
\LOGDCFL\ $\subseteq$ \SLOGCFL\ $\subseteq$ \LOGCFL.
\end{theorem}

\begin{theorem}\label{thm:ustreal}
\USTREAL\ is \SLOGCFL-complete.
\end{theorem}

\begin{corollary}\label{cor:ustrealnoepsilon}
\USTREAL\ with no $\epsilon$-edges is \SLOGCFL-complete.
\end{corollary}

\section{More Realizability Problems between \logspace\ and \LOGCFL}\label{sec:realproblems}

As noted earlier, an instance $\mathcal{G}$ of \STREAL\ is represented by an $n \times n$ standard matrix $\mathcal{E}$ and an $n^2 \times n^2$ gap matrix ${\Upsilon}$. The vertices of $\mathcal{G}$ are labeled with $\{\alpha_1, \dots, \alpha_k\}$. In this section, we define more graph realizability problems based on the symmetry of the gap and standard matrices and $\mathcal{E}$ and the number of distinct vertex labels (i.e., number of stack symbols, denoted by $k$). We define the corresponding complexity classes as the set of all languages that are logspace reducible to the corresponding graph realizability problem. Table 1 summarizes all the definitions. The prefix ${\bf{S}}$ is used to denote the symmetry of the standard matrix. The prefix ${\bf{SGS}}$ is used to denote the symmetry of the standard {\em{and}} gap matrices. A moment of thought would reveal that the case of symmetric gap matrix and asymmetric standard matrix does not make much sense. The prefix ${\bf{1}}$ is used to denote that there is only one vertex label.

\begin{table}\label{defnstbl}
\begin{tabular}{l*{6}{c}r}
    Complexity class & Number of stack symbols & Standard Matrix & Gap Matrix  \\
    \hline
    \LOGCFL\ & $k \geq 2$ & asymmetric & asymmetric  \\
    \SLOGCFL\ & $k \geq 2$ & symmetric  & asymmetric \\
    \SGSLOGCFL\ & $k \geq 2$ & symmetric & symmetric \\
    \oneLOGCFL\ & $k = 1$ & asymmetric & asymmetric \\
    \oneSLOGCFL\ & $k = 1$ & symmetric & asymmetric \\
    \oneSGSLOGCFL\ & $k = 1$ & symmetric & symmetric \\
\end{tabular}
\caption{Graph realizability problems between \logspace\ and \LOGCFL.}
\end{table}

\subsection{Realizability with Symmetric Gap}\label{subsec:sgslogcfl}

We are given an undirected graph $\mathcal{G}(V,E)$, a vertex labeling function $L_{V}:V{\rightarrow}\{\alpha_1,\alpha_2,\dots,\alpha_k\}$ and an edge labeling function $L_{E}:E{\rightarrow}\{push,pop,\epsilon\}$. The edge labels are ``symmetric" as defined in \secref{sec:ustconn}. The pair $(s,t)$, where $s,t \in V$, is said to be {\bf{realizable with symmetric gap}} if the following two conditions hold :
\begin{itemize}
\item{There is an undirected path (say $P$) from $s$ to $t$.}
\item{The concatenation of the vertex and edge labels along the path $P$ is a {\em{realizable string with symmetric gap}} (see \defref{defn:symgapreal}).}
\end{itemize}

\begin{definition}\label{defn:symgapreal}
Let $\mathcal{A} = \{push,pop,\epsilon,\alpha_1,\alpha_2,\dots,\alpha_k\}$ be the set of alphabets. A {\bf{realizable string with symmetric gap}} is a nonempty string of alphabets from $\mathcal{A}$, defined in the following recursive manner :
\begin{itemize}
\item{for all $1 \leq i \leq k$, ``$\alpha_i$" is a realizable string.}
\item{for all $1 \leq i \leq k$, ``${\alpha_i}\ \epsilon\ {\alpha_i}$" is a realizable string.}
\item{if $S$ is a realizable string then so is ``${\alpha_i}\ push\ S\ pop\ {\alpha_i}$", for all $1 \leq i \leq k$.}
\item{if $S$ is a realizable string then so is ``${\alpha_i}\ pop\ S\ push\ {\alpha_i}$", for all $1 \leq i \leq k$.}
\item{for all $1 \leq i \leq k$, if ``${\alpha_i}\ S_1\ {\alpha_i}$" and ``${\alpha_i}\ S_2\ {\alpha_i}$" are realizable strings then so is ``${\alpha_i}\ S_1\ {\alpha_i}\ S_2\ {\alpha_i}$".}
\end{itemize}
\end{definition}

Since the edge labels are symmetric, $(s,t)$ is realizable if and only if $(t,s)$ is realizable. We initialize the gap matrix as described in \secref{subsec:gapmatrixdefn}. By the definition of {\em{realizable string with symmetric gap}}, $(a{\leadsto}(c,d){\leadsto}b)$ if and only if $(c{\leadsto}(a,b){\leadsto}d)$. Hence the corresponding $n^2 \times n^2$ gap matrix ${\Upsilon}$ is a symmetric matrix. We denote this symmetry by $(a{\leftrightsquigarrow}(c,d){\leftrightsquigarrow}b)$.

\begin{framed}
\noindent \SGUSTREAL\ : Given an undirected graph $\mathcal{G}(V,E)$ with vertices labeled from $\{\alpha_1,\alpha_2,\dots,\alpha_k\}$ and {\em{symmetric}} edge labels from $\{push,pop,\epsilon\}$ and two distinguished nodes $s$ and $t$, decide if $s$ and $t$ are {\em{realizable with symmetric gap}} in $\mathcal{G}$.
\end{framed}

\begin{framed}
\noindent \SGSLOGCFL\ is the class of languages that are logspace reducible to \SGUSTREAL.
\end{framed}

\subsection{Realizability with one stack symbol}

The complexity classes \oneLOGCFL, \oneSLOGCFL\ and \oneSGSLOGCFL\ are obtained by restricting \LOGCFL, \SLOGCFL\ and \SGSLOGCFL\ respectively to use only one stack symbol i.e., by insisting that $k = 1$ in the above definitions. Since the vertices are all labeled with one label, we may omit the vertex labels in the definitions. After omitting the vertex labels, the corresponding {\em{realizability}} can be defined using a context-free language as shown below.

\subsubsection{\oneLOGCFL}\label{one-logcfl}

\oneLOGCFL\ is the class of languages that are logspace reducible to the following graph realizability problem. We are given a directed graph $\mathcal{G}(V,E)$, with edges labeled from $\{push,pop,\epsilon\}$. The ordered pair $(s,t)$, where $s,t \in V$, is said to be realizable if the following two conditions hold :
\begin{itemize}
\item{There is a directed path (say $P$) from $s$ to $t$.}
\item{The concatenation of the edge labels on the path $P$ is a string produced by the following context-free grammar :\ \ $S \rightarrow S\ S$;\ $S \rightarrow push\ S\ pop$;\ $S \rightarrow \epsilon$;\ $S \rightarrow \emptyset$. Here $\emptyset$ denotes the empty string.
}
\end{itemize}

\subsubsection{\oneSLOGCFL}\label{one-slogcfl}

We are given an undirected graph $\mathcal{G}(V,E)$, with the edges labeled from $\{push,pop,\epsilon\}$. Moreover, the edge labels are ``symmetric" as defined in \secref{sec:ustconn}. The pair $(s,t)$, where $s,t \in V$, is said to be {\em{realizable}} if there is an undirected path (say $P$) from $s$ to $t$ and the concatenation of the edge labels along the path $P$ is a string produced by the context-free grammar mentioned in \secref{one-logcfl}. Since the edge labels are symmetric, $(s,t)$ is realizable if and only if $(t,s)$ is realizable. \oneSLOGCFL\ is the class of languages that are logspace reducible to this undirected graph realizability problem.

\subsubsection{\oneSGSLOGCFL}\label{one-sgslogcfl}

\oneSGSLOGCFL\ is the class of languages that are logspace reducible to the following graph realizability problem. We are given an undirected graph $\mathcal{G}(V,E)$, with the edges labeled from $\{push,pop,\epsilon\}$. The edge labels are ``symmetric" as defined in \secref{sec:ustconn}. The pair $(s,t)$, where $s,t \in V$, is said to be realizable if the following two conditions hold :
\begin{itemize}
\item{There is a simple undirected path (say $P$) from $s$ to $t$.}
\item{The concatenation of the edge labels on the path $P$ is a string produced by the following context-free grammar :\ \ $S \rightarrow S\ S$;\ $S \rightarrow push\ S\ pop$;\ $S \rightarrow pop\ S\ push$;\ $S \rightarrow \epsilon$;\ $S \rightarrow \emptyset$. Here $\emptyset$ denotes the empty string.
}
\end{itemize}

\subsection{Relationship among the Realizability Problems}

By definition, we have the following inclusions : (i) \SGSLOGCFL\ $\subseteq$ \SLOGCFL\ $\subseteq$ \LOGCFL, (ii) \oneSGSLOGCFL\ $\subseteq$ \oneSLOGCFL\ $\subseteq$ \oneLOGCFL, (iii) \oneLOGCFL\ $\subseteq$ \LOGCFL, (iv) \oneSLOGCFL\ $\subseteq$ \SLOGCFL\ and (v) \oneSGSLOGCFL\ $\subseteq$ \SGSLOGCFL. Independent to our work, Allender and Lange \cite{allender-lange} defined symmetric AuxPDAs and proved that every language accepted by a nondeterministic auxiliary pushdown automaton in polynomial time can be accepted by a symmetric auxiliary pushdown automaton in polynomial time. Their definition of symmetric AuxPDAs is equivalent to ours \cite{allender-personal-comm}. Borodin et. al. \cite{BCDRT89} proved that \LOGCFL\ = \coLOGCFL. The following theorem and its corollary are immediate.

\begin{theorem}\label{thm:all-lange-theorem}
{\em{(Allender and Lange \cite{allender-lange})}}. \SLOGCFL\ = \LOGCFL.
\end{theorem}

\begin{corollary}
\SLOGCFL\ = \coSLOGCFL.
\end{corollary}

\subsection{Realizability Problems between \logspace\ and \NL}\label{subsec:balanced-defn}

All the realizability problems defined above are generalizations of \USTCONN. Hence, the corresponding complexity classes contain \logspace. We now prove that \NL\ = \oneLOGCFL. Hence, \logspace\ = \SL\ $\subseteq$ \oneSGSLOGCFL\ $\subseteq$ \oneSLOGCFL\ $\subseteq$ \oneLOGCFL\ = \NL. We introduce two natural graph connectivity problems characterizing \oneSGSLOGCFL\ and \oneSLOGCFL.

\begin{theorem}\label{thm:nl-1logcfl}
\NL\ = \oneLOGCFL.
\end{theorem}

\begin{corollary}
\logspace\ = \SL\ $\subseteq$ \oneSGSLOGCFL\ $\subseteq$ \oneSLOGCFL\ $\subseteq$ \oneLOGCFL\ = \NL.
\end{corollary}

Let $\mathcal{G}(V,E)$ be a directed graph. Let $\mathcal{G'}(V,E')$ be the underlying undirected graph of $\mathcal{G}$. Let $P$ be a path in $\mathcal{G'}$. Let $e = (u,v)$ be an edge along the path $P$. Edge $e$ is called {\em{neutral}} edge if both $(u,v)$ and $(v,u)$ are in $E$. Edge $e$ is called {\em{forward}} edge if $(u,v) \in E$ and $(v,u) \notin E$. Edge $e$ is called {\em{backward}} edge if $(u,v) \notin E$ and $(v,u) \in E$.

A path (say $P$) from $s \in V$ to $t \in V$ in $\mathcal{G'}(V,E')$ is called {\em{balanced}} if the number of forward edges along $P$ is equal to the number of backward edges along $P$. A balanced path might have any number of neutral edges. By definition, if there is a balanced path from $s$ to $t$ then there is a balanced path from $t$ to $s$. The path $P$ may not be a simple path. We are concerned with balanced paths of length at most $n$. See \secref{sec:bstconn} in the {\bf{appendix}} for more details and variants of balanced connectivity problems.

\begin{framed}
\noindent \BSTCONN\ : Given a directed graph $\mathcal{G}(V,E)$ and two distinguished nodes $s$ and $t$, decide if there is {\em{balanced}} path (of length at most $n$) between $s$ and $t$.
\end{framed}

Let $P$ be a path from $s \in V$ to $t \in V$ in $\mathcal{G}(V,E)$. We say $v \in P$ if the vertex $v$ is on the path $P$. For $v \in P$ we denote by $P_v$ the subpath of $P$ starting from $s$ and ending at $v$. We say that $P$ is {\em{positive}} if the number of forward edges of $P_v$ is at least the number of backward edges of $P_v$, for all $v \in P$. In other words, the number of forward edges minus the number of backward edges of $P_v$ is positive, for all $v \in P$. We say that $P$ is {\em{positive balanced}} if $P$ is positive and balanced. By definition, if there is a positive balanced path from $s$ to $t$ then there is a positive balanced path from $t$ to $s$.

\begin{framed}
\noindent \PBSTCONN\ : Given a directed graph $\mathcal{G}(V,E)$ and two distinguished nodes $s$ and $t$, decide if there is {\em{positive balanced}} path (of length at most $n$) between $s$ and $t$.
\end{framed}

\begin{theorem}\label{thm:1sgslogcfl}
\BSTCONN\ is \oneSGSLOGCFL-complete.
\end{theorem}

\begin{theorem}\label{thm:1slogcfl}
\PBSTCONN\ is \oneSLOGCFL-complete.
\end{theorem}

\begin{figure}[htp]
\begin{center}
\includegraphics[width=4in]{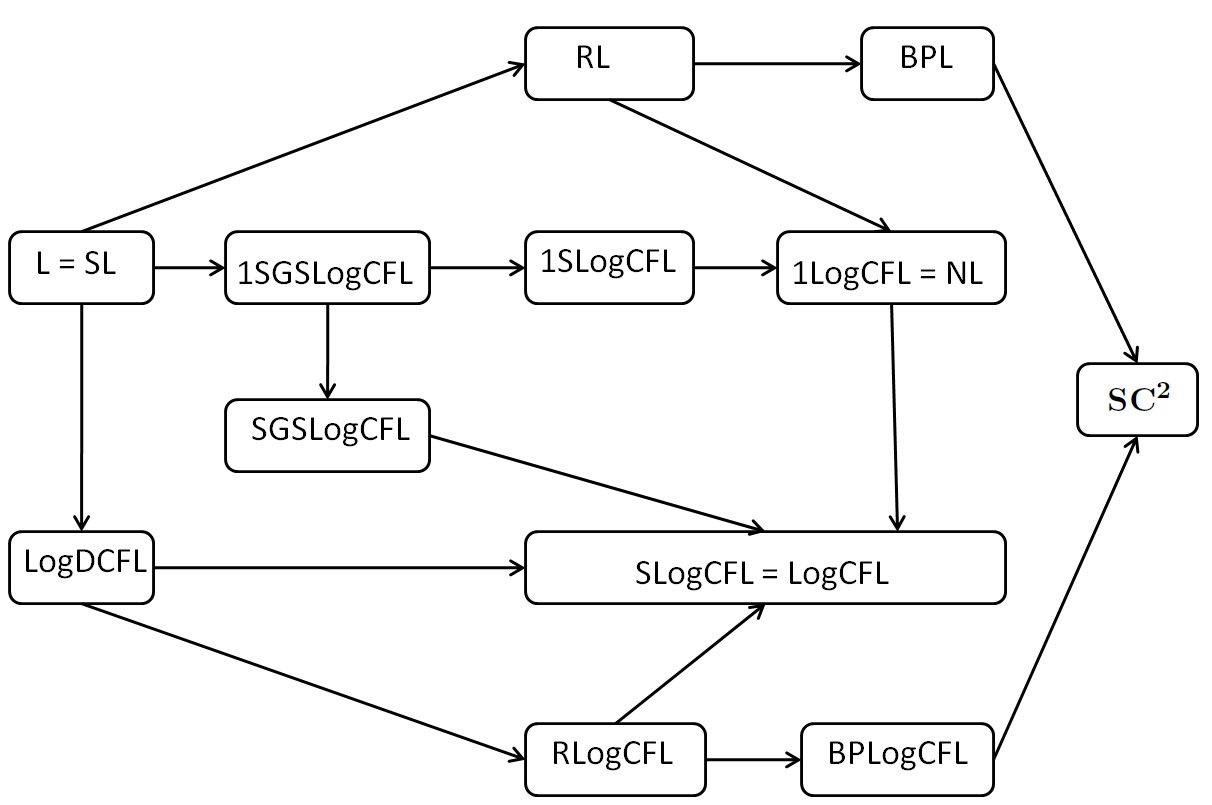}
\end{center}
\caption{Relationship among the complexity classes. A directed edge from class ${\bf{A}}$ to class ${\bf{B}}$ shows that ${\bf{A}} \subseteq {\bf{B}}$. In addition, \RL\ $\subseteq$ \RLOGCFL\ and \BPL\ $\subseteq$ \BPLOGCFL. \BSTCONN\ is \oneSGSLOGCFL-complete and \PBSTCONN\ is \oneSLOGCFL-complete.}
\label{fig:real-classes}
\end{figure}

Figure \ref{fig:real-classes} summarizes the relationship among the above defined classes.

\section{Transitive Closure}\label{sec:transitive}

The definitions and theorems in this section apply to all the graph realizability problems defined above. We present the definitions and theorems for \STREAL, the most general graph realizability problem.

\begin{definition}
Let $\mathcal{G} = {\langle}{\Upsilon},\mathcal{E}{\rangle}$ be an instance of \STREAL. The {\bf{transitive closure}} of $\mathcal{G}$, denoted by $\mathcal{G}^* = {\langle}{\Upsilon}^*,\mathcal{E}^*{\rangle}$, is a pair of gap and standard matrix such that for all $a,b,c,d \in V$, \\
\indent (i) $\mathcal{E}^*[a][b] = 1$ iff $(a{\leadsto}b)$ and \\
\indent (ii) ${\Upsilon}^*[a,(c,d),b] = 1$ iff $(a,b)$ is realizable with gap $(c,d)$.
\end{definition}

\subsection{Tensor Products}\label{subsec:tensor}

We now present several tensor products acting on $\mathcal{E}$ and ${\Upsilon}$. The products $\otimes_1$ to $\otimes_5$ are introduced in \cite{venkat-auxpda}. We introduce $\otimes_6$ and $\otimes_7$. These products update the standard matrix $\mathcal{E}$ and the gap matrix $\Upsilon$ with new ``connectivity information" of $\mathcal{G}$. Let $\mathcal{E}$, $\mathcal{E}_1$, $\mathcal{E}_2$ represent standard matrices and ${\Upsilon}$, ${\Upsilon}_1$, ${\Upsilon}_2$ represent gap matrices. Let $a,b,c,d,z$ represent the vertices of $\mathcal{G}$. Matrices indexed by two (resp. four) indices are standard (resp. gap) matrices. When we are dealing with boolean matrices, all the summations (resp. multiplications) are interpreted as boolean $\vee$ (resp. boolean $\wedge$).

\begin{enumerate}

\item{
If $(a{\leadsto}z)$ and $(z{\leadsto}b)$ then $(a{\leadsto}b)$ :
\begin{center}
$(\mathcal{E}_1 \otimes_1 \mathcal{E}_2)[a,b] = \displaystyle\sum_z {\mathcal{E}_1}[a,z]{\cdot}{\mathcal{E}_2}[z,b]$.
\end{center}
}

\item{
If $(a{\leadsto}(c,d){\leadsto}b)$ and $(c{\leadsto}d)$ then $(a{\leadsto}b)$ :
\begin{center}
$({\Upsilon} \otimes_2 {\mathcal{E}})[a,b] = \displaystyle\sum_{c,d} {\Upsilon}[a,(c,d),b]{\cdot}{\mathcal{E}}[c,d]$.
\end{center}
}

\item{
If $(a{\leadsto}(c,d){\leadsto}b)$ and $(b{\leadsto}z)$ then $(a{\leadsto}(c,d){\leadsto}z)$ :
\begin{center}
$({\Upsilon} \otimes_3 {\mathcal{E}})[a,(c,d),z] = \displaystyle\sum_b {\Upsilon}[a,(c,d),b]{\cdot}{\mathcal{E}}[b,z]$.
\end{center}
}

\item{
If $(z{\leadsto}a)$ and $(a{\leadsto}(c,d){\leadsto}b)$ then $(z{\leadsto}(c,d){\leadsto}b)$ :
\begin{center}
$({\mathcal{E}} \otimes_4 {\Upsilon})[z,(c,d),b] = \displaystyle\sum_a {\mathcal{E}}[z,a]{\cdot}{\Upsilon}[a,(c,d),b]$.
\end{center}
}

\item{
If $(a{\leadsto}(c,d){\leadsto}b)$ and $(c{\leadsto}(e,f){\leadsto}d)$ then $(a{\leadsto}(e,f){\leadsto}b)$ :
\begin{center}
$({{\Upsilon}_1} \otimes_5 {{\Upsilon}_2})[a,(e,f),b] = \displaystyle\sum_{c,d} {{\Upsilon}_1}[a,(c,d),b]{\cdot}{{\Upsilon}_2}[c,(e,f),d]$.
\end{center}
}

\item{
If $(a{\leadsto}(c,d){\leadsto}b)$ and $(z{\leadsto}d)$ then $(a{\leadsto}(c,z){\leadsto}b)$ :
\begin{center}
$({\Upsilon} \otimes_6 {\mathcal{E}})[a,(c,z),b] = \displaystyle\sum_{d} {\Upsilon}[a,(c,d),b]{\cdot}{\mathcal{E}}[z,d]$.
\end{center}
}

\item{
If $(a{\leadsto}(c,d){\leadsto}b)$ and $(c{\leadsto}z)$ then $(a{\leadsto}(z,d){\leadsto}b)$ :
\begin{center}
$({\Upsilon} \otimes_7 {\mathcal{E}})[a,(z,d),b] = \displaystyle\sum_{c} {\Upsilon}[a,(c,d),b]{\cdot}{\mathcal{E}}[c,z]$.
\end{center}
}

\end{enumerate}

\subsection{Computing Transitive Closure}\label{subsec:ckt-square}

Given $\mathcal{G} = {\langle}{\Upsilon},\mathcal{E}{\rangle}$ the following algorithm computes ${\bf{Square}}(\mathcal{G})$. This algorithm is based on a parsimonious simulation of \LOGCFL\ by \SACone\ circuits given by Niedermeier and Rossmanith \cite{NiedermeierR95}. Implementation of ${\bf{Square}}({\langle}{\Upsilon},\mathcal{E}{\rangle})$ using the above mentioned tensor products is shown below. \thref{thm:ckt-square} implies a natural polynomial time algorithm to solve \STREAL.

\noindent \line(1,0){450} \\
\noindent ${\bf{Square}}({\langle}{\Upsilon},\mathcal{E}{\rangle})$ \\

\indent for all $a,b \in V$ update $\mathcal{E}$ as follows :
\begin{eqnarray*}
\mathcal{E}[a,b] & = & \displaystyle\sum_{c,e,f,g,d}{\Upsilon}[a,(c,d),b]{\cdot}{\Upsilon}[c,(e,f),g]{\cdot}\mathcal{E}[e,f]{\cdot}\mathcal{E}[g,d] \\
\end{eqnarray*}
\indent for all $a,b,c,d \in V$ update ${\Upsilon}$ as follows :
\begin{eqnarray*}
{\Upsilon}[a,(c,d),b] & = & \displaystyle\sum_{c',e',f',g',d'}{\Upsilon}[a,(c'd'),b]{\cdot}{\Upsilon}[c',(e',f'),g']{\cdot}{\Upsilon}[e',(c,d),f']{\cdot}\mathcal{E}[g',d'] \\
& + & \displaystyle\sum_{c',e',f',g',d'}{\Upsilon}[a,(c'd'),b]{\cdot}{\Upsilon}[c',(e',f'),g']{\cdot}\mathcal{E}[e',f']{\cdot}{\Upsilon}[g',(c,d),d'] \\
\end{eqnarray*}
\indent {\bf{return}} ${\langle}{\Upsilon},\mathcal{E}{\rangle}$ \\
\noindent \line(1,0){450} \\

\noindent \line(1,0){300} \\
\noindent ${\bf{Square}}({\langle}{\Upsilon},\mathcal{E}{\rangle})$ \\
\indent $\mathcal{E} = (\Upsilon \otimes_2((\Upsilon \otimes_2 \mathcal{E})\otimes_1 \mathcal{E}))$ \\
\indent $\Upsilon = (\Upsilon \otimes_5((\Upsilon \otimes_5 \Upsilon)\otimes_3 \mathcal{E})) 
                         + (\Upsilon \otimes_5((\Upsilon \otimes_2 \mathcal{E})\otimes_4 \Upsilon)) $\\
\indent {\bf{return}} ${\langle}{\Upsilon},\mathcal{E}{\rangle}$ \\
\noindent \line(1,0){300} \\

\begin{theorem}\label{thm:ckt-square}
Let $\mathcal{G}$ be an instance of \STREAL. $\mathcal{G}^* = {\langle}{\Upsilon}^*,\mathcal{E}^*{\rangle}$ can be computed using $O({\log}n)$ repeated applications of ${\bf{Square}}(\mathcal{G})$.
\end{theorem}

\subsection{Simple Squaring Operation}

The following algorithm ${\bf{SimpleSquare}}$ is a more intuitive squaring operation. It plays a crucial role in the proofs of correctness of parallel and space efficient algorithms for \SGSLOGCFL\ (see \secref{sec:parallel-sgslogcfl} and \secref{sec:sgslogcfl-logloglog}).

\noindent \line(1,0){200} \\
\noindent ${\bf{SimpleSquare}}({\langle}{\Upsilon},\mathcal{E}{\rangle})$ \\
\indent        $\mathcal{E} = \mathcal{E} \otimes_1 \mathcal{E}$ \\
\indent        $\mathcal{E} = \Upsilon \otimes_2 \mathcal{E}$ \\
\indent        $\Upsilon = \Upsilon \otimes_3 \mathcal{E}$ \\
\indent        $\Upsilon = \mathcal{E} \otimes_4 \Upsilon$ \\
\indent        $\Upsilon = \Upsilon \otimes_5 \Upsilon$ \\
\indent        $\Upsilon = \Upsilon \otimes_6 \mathcal{E}$ \\
\indent        $\Upsilon = \Upsilon \otimes_7 \mathcal{E}$ \\
{\bf{return}} ${\langle}{\Upsilon},\mathcal{E}{\rangle}$ \\
\noindent \line(1,0){200}

\begin{theorem}\label{thm:simple-square}
Let $\mathcal{G}$ be an instance of \STREAL. $\mathcal{G}^* = {\langle}{\Upsilon}^*,\mathcal{E}^*{\rangle}$ can be computed using $O({\log}n)$ repeated applications of ${\bf{SimpleSquare}}(\mathcal{G})$.
\end{theorem}

\section{Parallel algorithms for \SGSLOGCFL}\label{sec:parallel-sgslogcfl}

Let $\mathcal{G} = {\langle}{\Upsilon},\mathcal{E}{\rangle}$ be an instance of \SGUSTREAL. Let the vertices of $\mathcal{G}$ be $V = \{1,2,\dots,n\}$. $\mathcal{G}$ is represented by an $n \times n$ standard matrix $\mathcal{E}$ and an $n^2 \times n^2$ gap matrix ${\Upsilon}$. In this section, we present parallel algorithms to compute $\mathcal{G}$'s transitive closure $\mathcal{G}^* = {\langle}{\Upsilon}^*,\mathcal{E}^*{\rangle}$. Let $V^2 = V \times V$ be the set of pairs of vertices. In the rest of this paper the term ``vertex" refers to elements from $V$ as well as $V^2$. Let $V^4 = V \times V \times V \times V$. $\mathcal{G}$ has two types of edges. The standard edges from $V^2$ are present in $\mathcal{E}$ and the gap edges from $V^4$ are present in ${\Upsilon}$. In the rest of this paper the term ``edge" refers to elements from $V^2$ as well as $V^4$.

\begin{definition}
A subset of vertices $S \subseteq V$ is a {\bf{standard component}} ($s$-component) of $\mathcal{G}$ iff for all $u,v \in S$ it holds that $(u{\leadsto}v)$ and $(v{\leadsto}u)$.
\end{definition}

\begin{definition}
A subset $S \subseteq V^2$ is a {\bf{gap component}} ($g$-component) of $\mathcal{G}$ iff for all $(a,b),(c,d) \in S$ it holds that $(a{\leadsto}(c,d){\leadsto}b)$ and $(c{\leadsto}(a,b){\leadsto}d)$.
\end{definition}

In the rest of this paper the term ``component" refers to both standard and gap components. If there is ambiguity we will explicitly say $s$-component or $g$-component.

A {\em{pseudotree}} $P = (C,D)$ is a maximal connected directed graph with $|C| = k$ vertices and $|D| = k$ arcs for some $k$, for which each vertex has outdegree one. Note that every pseudotree has exactly one simple directed cycle (which may be a self-loop). The number of arcs in the cycle of a pseudoree $P$ is its {\em{circumference}}. A {\em{rooted tree}} is a pseudotree whose cycle is a self-loop on some vertex $r$ called the {\em{root}}. A {\em{rooted star}} $R$ with root $r$, is a rooted tree whose arcs are of the form $(x,r)$ with $x \in R$. A {\em{pseudoforest}} is a collection of pseudotrees. \\

\noindent {\bf{Symmetric Squaring}} : We first present a simplified squaring algorithm when the input graph is an instance of \SGUSTREAL. Here the matrices $\mathcal{E}$ and ${\Upsilon}$ are symmetric i.e., $\mathcal{E}[a,b] = \mathcal{E}[b,a]$ and ${\Upsilon}[(a,b),(c,d)] = {\Upsilon}[(c,d),(a,b)]$. Moreover, ${\Upsilon}[(a,b),(c,d)] = {\Upsilon}[(a,b),(d,c)] = {\Upsilon}[(b,a),(c,d)] = {\Upsilon}[(b,a),(d,c)]$. Due to this symmetry, the products $\otimes_3$, $\otimes_4$, $\otimes_6$ and $\otimes_7$ are equivalent. \corref{cor:symmetric-square} follows from \thref{thm:simple-square}.

\noindent \line(1,0){200} \\
\noindent ${\bf{SymmetricSquare}}({\langle}{\Upsilon},\mathcal{E}{\rangle})$ \\
\indent        $\mathcal{E} = \mathcal{E} \otimes_1 \mathcal{E}$ \\
\indent        $\mathcal{E} = \Upsilon \otimes_2 \mathcal{E}$ \\
\indent        $\Upsilon = \Upsilon \otimes_3 \mathcal{E}$ \\
\indent        $\Upsilon = \Upsilon \otimes_5 \Upsilon$ \\
{\bf{return}} ${\langle}{\Upsilon},\mathcal{E}{\rangle}$ \\
\noindent \line(1,0){200}

\begin{corollary}\label{cor:symmetric-square}
Let $\mathcal{G}$ be an instance of \SGUSTREAL. $\mathcal{G}^*$ can be computed using $O({\log}n)$ repeated applications of ${\bf{SymmetricSquare}}(\mathcal{G})$.
\end{corollary}

\subsection{An $O({\log}^2n)$ time parallel algorithm}

\begin{algorithm}{{\bf{Connect}}($\mathcal{G} = {\langle}{\Upsilon},\mathcal{E}{\rangle}$)}
\begin{algorithmic}[1]
\STATE ${\mathcal{E}}^* \leftarrow {\mathcal{E}}$
\STATE ${\Upsilon}^* \leftarrow {\Upsilon}$

\STATE {\bf{for all}} $i$ {\bf{do}} $X_{\mathcal{E}}(i) = i$
\STATE {\bf{for all}} $i$ {\bf{do}} $X_{\Upsilon}(i,j) = (i,j)$
\vspace{0.1in}

\FOR{$O({\log}{n})$ iterations}
\vspace{0.1in}

\STATE {\bf{for all}} $i$ {\bf{do}} $Temp_{\mathcal{E}}(i) \leftarrow StandardHook(i)$
\STATE {\bf{for all}} $i$ {\bf{do}} $Temp_{\mathcal{E}}(i) \leftarrow \mbox{min}_j\{Temp_{\mathcal{E}}(j)\ |\ X_{\mathcal{E}}(j) = i\ \mbox{and}\ Temp_{\mathcal{E}}(j) \neq i\}$
\STATE \indent if none then $Temp_{\mathcal{E}}(i) \leftarrow X_{\mathcal{E}}(i)$ \\
\vspace{0.1in}

\STATE {\bf{for all}} $i$ {\bf{do}} $Temp_{\Upsilon}(i,j) \leftarrow GapHook(i,j)$
\STATE {\bf{for all}} $i$ {\bf{do}} $Temp_{\Upsilon}(i,j) \leftarrow \mbox{min}_{(k,l)}\{Temp_{\Upsilon}(k,l)\ |\ X_{\Upsilon}(k,l) = (i,j)\ \mbox{and}\ Temp_{\Upsilon}(k,l) \neq (i,j)\}$
\STATE \indent if none then $Temp_{\Upsilon}(i,j) \leftarrow X_{\Upsilon}(i,j)$
\vspace{0.1in}

\STATE {\bf{for all}} $i$ {\bf{do}} $X_{\mathcal{E}}(i) \leftarrow Temp_{\mathcal{E}}(i)$
\STATE {\bf{for all}} $(i,j)$ {\bf{do}} $X_{\Upsilon}(i,j) \leftarrow Temp_{\Upsilon}(i,j)$
\vspace{0.1in}

\FOR{$O({\log}{n})$ iterations}
\STATE {\bf{for all}} $i$ {\bf{do}} $Temp_{\mathcal{E}}(i) \leftarrow Temp_{\mathcal{E}}(Temp_{\mathcal{E}}(i))$
\STATE {\bf{for all}} $(i,j)$ {\bf{do}} $Temp_{\Upsilon}(i,j) \leftarrow Temp_{\Upsilon}(Temp_{\Upsilon}(i,j))$
\ENDFOR
\vspace{0.1in}

\STATE {\bf{for all}} $i$ {\bf{do}} $X_{\mathcal{E}}(i) \leftarrow \mbox{min}\{Temp_{\mathcal{E}}(i), X_{\mathcal{E}}(Temp_{\mathcal{E}}(i))\}$
\STATE {\bf{for all}} $(i,j)$ {\bf{do}} $X_{\Upsilon}(i,j) \leftarrow \mbox{min}\{Temp_{\Upsilon}(i,j), X_{\Upsilon}(Temp_{\Upsilon}(i,j))\}$
\vspace{0.1in}

\STATE {\bf{for all}} $i,j$ {\bf{do}} {\bf{if}} $X_{\mathcal{E}}(i) = X_{\mathcal{E}}(j)$ {\bf{then}} ${\mathcal{E}}^*[i,j] \leftarrow 1$.
\STATE {\bf{for all}} $i,j,k,l$ {\bf{do}} {\bf{if}} $X_{\Upsilon}(i,j) = X_{\Upsilon}(k,l)$ {\bf{then}} ${\Upsilon}^*[i,(k,l),j] \leftarrow 1$.
\vspace{0.1in}

\ENDFOR
\vspace{0.1in}

\RETURN $\mathcal{G}^* = {\langle}{\Upsilon}^*,\mathcal{E}^*{\rangle}$
\end{algorithmic}
\end{algorithm}

\begin{algorithm}{{\bf{StandardHook}}($i$)}
\begin{algorithmic}[1]
\STATE $S_1 \leftarrow \{X_{\mathcal{E}}(j)\ |\ \mathcal{E}^*[i,j] = 1\ \mbox{and}\ X_{\mathcal{E}}(j) \neq X_{\mathcal{E}}(i)\}$
\STATE $S_2 \leftarrow \{X_{\mathcal{E}}(j)\ |\ \Upsilon^*[i,(k,k),j] = 1\ \mbox{and}\ X_{\mathcal{E}}(j) \neq X_{\mathcal{E}}(i)\}$
\STATE $S = S_1 \cup S_2$
\IF {$S = \emptyset$}
\RETURN $X_{\mathcal{E}}(i)$
\ELSE
\RETURN min($S$)
\ENDIF
\end{algorithmic}
\end{algorithm}

\begin{algorithm}{{\bf{GapHook}}($i,j$)}
\begin{algorithmic}[1]
\STATE $S_1 \leftarrow \{X_{\Upsilon}(k,l)\ |\ \Upsilon^*[i,(k,l),j] = 1\ \mbox{and}\ X_{\Upsilon}(k,l) \neq X_{\mathcal{E}}(i,j)\}$
\STATE $S_2 \leftarrow \{X_{\Upsilon}(k,j)\ |\ \mathcal{E}^*[i,k] = 1\ \mbox{and}\ X_{\Upsilon}(k,j) \neq X_{\Upsilon}(i,j)\}$
\STATE $S = S_1 \cup S_2$
\IF {$S = \emptyset$}
\RETURN $X_{\Upsilon}(i,j)$
\ELSE
\RETURN min($S$)
\ENDIF
\end{algorithmic}
\end{algorithm}

We will assume that there is one processor $P_i$ assigned to each vertex $i \in V$, one processor $P_{ij}$ assigned to each edge $(i,j) \in V^2$ and one processor $P_{ijkl}$ assigned to each {\em{gap edge}} $(i,j,k,l) \in V^4$. We use a vector $X_{\mathcal{E}}$ of length $n$ to specify the $s$-components of $\mathcal{G}$ as follows : if $V_c \subseteq V$ is any $s$-component, then for all $i \in V_c$, $X_{\mathcal{E}}(i)$ equals the least element of $V_c$. We use an $n \times n$ matrix $X_{\Upsilon}$ to specify the $g$-components of $\mathcal{G}$ as follows : if $W_c \subseteq V^2$ is any $g$-component, then for all $(i,j) \in W_c$, $X_{\Upsilon}(i,j)$ equals the lexicographically least element of $W_c$.

The algorithm {\bf{Connect}} iteratively computes the vectors $X_{\mathcal{E}}$ and $X_{\Upsilon}$ from the input $\mathcal{G} = {\langle}{\Upsilon},\mathcal{E}{\rangle}$ and updates ${\Upsilon}^*$ and $\mathcal{E}^*$. It is based on a hook and contract algorithm \cite{parallel-ustconn-log2n} that works as follows. The algorithm deals with ``components", which are sets of ``vertices" {\em{found}} to belong to the same (standard or gap) component of $\mathcal{G}$. Each component is equipped with an edge-list, a linked list of edges that connect it to other components. Initially each element from $V$ is an $s$-component by itself. Their edge-lists correspond to the undirected edges of $\mathcal{E}$. These components will eventually grow and become the corresponding $s$-components. Initially each element from $V^2$ is a $g$-component by itself. Their edge-lists correspond to the undirected edges of $\Upsilon$. These components will eventually grow and become the corresponding $g$-components. The algorithm proceeds as follows : \\

\noindent {\bf{repeat}} until there are no edges left :

\begin{enumerate}

\item{Each component picks an edge pointing to a lexicographically minimum vertex from its edge-list leading to a neighboring component and hooks by pointing to it. If a component has an empty edge-list, it hooks to itself. The details of hooking are presented in {\bf{StandardHook}} and {\bf{GapHook}}. Note that both these hooking steps use the previously computed connectivity information from {\em{both}} ${\Upsilon}^*$ and $\mathcal{E}^*$. These hooking processes create clusters of components called pseudotrees. The $s$-components form pseudotrees on the vertex set $V$ and $g$-components form pseudotrees on the vertex set $V^2$.}

\item{Each pseudotree is identified as a new component with one of its vertices as its representative. Each representative receives into its edge-list all the edges contained in the edge-lists of its pseudotree. At this stage the matrices $\mathcal{E}^*$ and ${\Upsilon}^*$ are updated with ``new" edges.}

\item{Edges internal to components are removed implicitly.}

\end{enumerate}

During the first iteration the edges connecting each vertex to neighboring vertices are examined (steps 6-11), and sets of vertices which are known to be connected are identified (steps 14-17). In effect, each such set of vertices is merged into a ``supervertex" which are specified by the vectors $X_{\mathcal{E}}(i)$ and $X_{\Upsilon}(i,j)$. For each $i$ in a supervertex, $X_{\mathcal{E}}(i)$ equals the smallest-numbered vertex in the supervertex. For each $(i,j)$ in a supervertex, $X_{\Upsilon}(i,j)$ equals the lexicographically first vertex in the supervertex. In succeeding iterations, the edges connecting each supervertex to neighboring supervertices are examined in steps 6-11, and sets of supervertices are merged in steps 14-17. The process continues until all the vertices in a (standard and gap) component have been merged into one gigantic supervertex.

\begin{theorem}\label{thm:parallel-log2n}
The algorithm {\bf{Connect}} finds $\mathcal{G}^* = {\langle}{\Upsilon}^*,\mathcal{E}^*{\rangle}$ in parallel time $O({\log}^2n)$ using $n^4$ processors in the CREW PRAM model.
\end{theorem}

{\bf{Connect}} algorithm is a generalization of the parallel algorithm presented in \cite{parallel-ustconn-log2n}. We added two hooking procedures (one for growing $s$-components and one for growing $g$-components). Unlike \cite{parallel-ustconn-log2n} the new edges found after the contraction step are added in the matrices ${\Upsilon}^*$ and $\mathcal{E}^*$ {\em{before}} starting the next hooking step.

The algorithms of \cite{parallel-ustconn-log3by2n} and \cite{parallel-ustconn-lognloglogn} can similarly be generalized to compute $\mathcal{G}^* = {\langle}{\Upsilon}^*,\mathcal{E}^*{\rangle}$ in parallel time $O({\log}^{3/2}n)$ and $O({\log}{n}{\log}{\log}{n})$ respectively. The processor bounds in all these algorithms is polynomial in $n$, the number of vertices of $\mathcal{G}$. We now present an outline of the parallel algorithms of \cite{parallel-ustconn-log3by2n} and \cite{parallel-ustconn-lognloglogn} and the necessary modifications to apply them to \SGSLOGCFL. We refer the reader to \cite{parallel-ustconn-log3by2n} and \cite{parallel-ustconn-lognloglogn} for low-level implementation details of these algorithms.

\subsection{An $O({\log}^{3/2}n)$ time parallel algorithm}

In the algorithm presented in the previous section the size of the components formed after hooking phase may vary a lot. A slow growing component may consist of as few as two vertices, whereas a fast-growing component may have as many as $n$ vertices for an $s$-component and $n^2$ vertices for a $g$-component. As a result the contraction (steps 14-17) requires $\Theta({\log}n)$ time in order to allow the biggest component to contract to a single vertex. The algorithm must iterate ${\log}n$ times so that a slow-growing component, which may only double its size in each iteration, can eventually grow to its full size. A crucial observation of \cite{parallel-ustconn-log3by2n} is that slow-growing components need little time to contract and fast-growing components require fewer iterations to grow to their full size.

Johnson and Metaxas \cite{parallel-ustconn-log3by2n} presented an algorithm in which components are scheduled to hook and contract according to their growth rate. Their algorithm schedules every component to grow by a factor of at least $2^{\sqrt{{\log}n}}$ in a phase of $O({\log}n)$ time. Hence, ${\sqrt{{\log}n}}$ phases suffice to find all connected components in the graph, for a total of $O({\log}^{3/2}n)$ time. Within a phase slow-growing components are scheduled to hook and contract in $o({\log}n)$ time repeatedly until they catch up with fast-growing components. Fast-growing components are left idle once they have achieved the intended size.

\begin{itemize}
\item{In the algorithm of \cite{parallel-ustconn-log2n} the vertices hook to a lexicographically minimum vertex. In Johnson-Metaxas algorithm vertices hook to the {\em{first edge}} in their edge-list. This creates pseudotrees of arbitrary circumference i.e., pseudotrees can have large cycles which are to be contracted properly in the contraction phase. Since exclusive writing is required, the usual pointer doubling technique will not terminate when applied to a cycle. Johnson and Metaxas \cite{parallel-ustconn-log3by2n} introduced {\em{cycle-reducing shortcutting}} technique to solve this problem. This technique (i) contracts a pseudotree into a rooted tree in time logarithmic in its circumference, (ii) contracts a rooted tree into a rooted star in time logarithmic in the length of its longest path.}

\item{It is expensive to compute the set of edges of all the components in a pseudotree without concurrent writing. Potentially there are a large number of components that hook together in the first step and therefore a large number of components that are ready to give their edge-lists simultaneously to the new super-component's edge-list. Johnson and Metaxas \cite{parallel-ustconn-log3by2n} introduced {\em{edge-plugging scheme}} which achieves the objective in constant time, irrespective of whether the component is yet contracted to a rooted star.}

\item{It is also expensive to have a component pick a mate. There may be a large number of edges internal to the component. The number of such edges grows every time components hook. These internal edges cannot be used to find a mate. Hence, a component may attempt to find a mate several times and will be unsuccessful if it picks an internal edge. Removing all the internal edges before picking an edge may also take a lot of time. Johnson and Metaxas \cite{parallel-ustconn-log3by2n} introduced a {\em{growth-control schedule}}. Components grow in size in a uniform way that controls their minimum sizes as long as continued growth is possible. The internal edges are identified and removed periodically to make hooking more efficient. The algorithm recognizes whether a component is growing too fast and therefore can be ignored.}
\end{itemize}

For implementation details of the above algorithm see \cite{parallel-ustconn-log3by2n}. As mentioned earlier, to get the corresponding parallel algorithm for \SGSLOGCFL\ we add two hooking procedures (one for growing $s$-components and one for growing $g$-components). After each contraction step the newly found edges are added in the matrices ${\Upsilon}^*$ and $\mathcal{E}^*$ .

\subsection{An $O({\log}{n}{\log}{\log}{n})$ time parallel algorithm}

The Chong Lam algorithm \cite{parallel-ustconn-lognloglogn} is also based on a hook and contract approach. The hooking process uses an ordering $<_d$ of the vertices such that $u <_d v$ iff the degree of $u$ is less than the degree of $v$ (or) the degrees are the same, but $u$ is less than $v$ in their lexicographic ordering. Before every phase, every vertex of the current {\em{supergraph}} is either active, inactive or done. All active and inactive vertices have nonzero degree, the done vertices have zero degree, and there are no multiedges between active vertices; the inactive vertices are organized in a set of hooking trees. Initially all vertices with nonzero degree are active, and the rest are done.

To choose their hooking edges, the active vertices of the graph perform the following steps in parallel. (i) if a vertex $v$ has a neighbor larger according to $<_d$ than itself, then $v$ hooks to the {\em{largest}} such neighbor. (ii) if after the first step all neighbors of $v$ are hooked to it, then $v$ hooks to itself. Otherwise, if after the first step a neighbor $u$ of $v$ is hooked to a vertex different from $v$, then $v$ hooks to $u$. This type of hooking scheme guarantees that any tree with a large degree must also contain a large number of vertices. The hooking schemes of \cite{parallel-ustconn-log2n, parallel-ustconn-log3by2n} suffer from creating pseudotrees with few vertices but a large degree.

Some of the current hooking trees are contracted to a representative vertex in a contraction phase. The representative vertex is the only vertex in the tree which is hooked to itself. Whether a tree is contracted is determined by a parameter. This parameter depends on the phase and sets an upper bound on the sum of the degrees of the vertices of the trees which are contracted. For every contracted tree, its representative becomes a new active vertex and the rest of its vertices become done. All multiedges between new active vertices are removed. The vertices of every uncontracted tree become inactive.

The processing required by a hooking phase is performed in parallel time $O({\log}d)$, where $d$ is the degree of the active vertex, using pointer jumping. Checking the degree of a hooking tree during the contraction phase is done in parallel time $O({\log}c)$, where $c$ is the contraction parameter, by using pointer jumping and a constant time edge-list plugging technique. \\

\noindent \line(1,0){200} \\
\noindent {{\bf{Connect}}($k$)} \\
\indent {\bf{MaxHook}}; \\
\indent {\bf{if}} $k > 0$ {\bf{then}} \\
\indent \indent {\bf{Connect}}($2^{2^k}$) \\
\indent \indent {\bf{Connect}}($k-1$) \\
\indent \indent {\bf{Connect}}($k-1$) \\
\indent {\bf{Contract}}($2^{2^{k+1}}$) \\
\noindent \line(1,0){200} \\

A call to {\bf{Connect}}(${\lceil}{{\log}{\log}n}{\rceil}$) contracts every connected component of the graph to a single vertex and all the other vertices are organized in a set of rooted parent trees such that the root of the tree of a vertex $u$ is the vertex to which the connected component of $u$ is contracted.

To generalize this algorithm to \SGSLOGCFL, we make the following modifications : (i) add two hooking procedures (one for growing $s$-components and one for growing $g$-components) (ii) the new edges found after every call {\em{Contract}} are added in the matrices ${\Upsilon}^*$ and $\mathcal{E}^*$ and the new degrees of the vertices are recomputed. The correctness of the algorithm follows by using \corref{cor:symmetric-square} in the correctness argument of \cite{parallel-ustconn-lognloglogn}, implying an $O({\log}{n}{\log}{\log}{n})$ time EREW parallel algorithm computing $\mathcal{G}^* = {\langle}{\Upsilon}^*,\mathcal{E}^*{\rangle}$.

\section{\SGSLOGCFL\ $\subseteq$ $DSPACE({\log}{n}{\log}{\log}{n})$}\label{sec:sgslogcfl-logloglog}

Trifonov's algorithm \cite{trifonov-logloglog} is based on the $O({\log}{n}{\log}{\log}{n})$ time deterministic EREW PRAM algorithm with $O(m + n)$ processors of Chong and Lam \cite{parallel-ustconn-lognloglogn} outlined in the previous section. This parallel algorithm is first simulated sequentially in linear space. Using this sequential algorithm a mathematical structure called {\em{configuration}} is defined. This configuration corresponds to the state of the sequential algorithm at a certain point of its execution. An ordering on the edges incident to a vertex is fixed, and the hooking is done sequentially for all active vertices. Using the sequence of configurations an $O({\log}^2{n})$ space algorithm, which instead of storing all of its current state recomputes parts of it when it needs them. This algorithm works pretty much like Savitch's algorithm \cite{Savitch70}.

The max-degree hooking scheme of \cite{parallel-ustconn-lognloglogn} ensures that small trees have small neighborhoods. Using the exploration walks on trees defined by Koucky ́ \cite{koucky-uts}, the levels of recursion of \cite{parallel-ustconn-lognloglogn} are implemented so that they process small trees in $o({\log}n)$ space. These walks essentially play the role of the edge-list plugging technique and pointer jumping techniques employed by the Chong-Lam algorithm. They allow us to traverse the pseudotrees space-efficiently.

The $O({\log}n)$ space per level is mainly due to storing vertices in the local variables of the functions, since each vertex takes $\Theta({\log}n)$ space. To overcome this bottleneck the functions are redefined so that they never keep a vertex in their local variables. The vertex $v$ is removed from the argument list of the functions. Instead of this argument, one current vertex is maintained in a global variable. All functions are programmed to return some ``information" about this vertex. A function which otherwise must return a vertex is defined so that after its execution the current vertex is its result. If needed the calling function keeps enough information locally to restore the original current vertex. The crucial part of the optimization is to avoid storing vertices locally and be able to move the current vertex temporarily, perform something at the new current vertex, and then return to the original current vertex. Instead of this going back and forth between the two vertices, using the reversibility of the moves along the edges and the exploration walks on the trees, the comparison is performed bit by bit. Aside from the information stored for	the	ways back, this takes only the $\Theta({\log}n{\log}{\log}n)$ space necessary to store the index of a bit. In this way the bottleneck of $\Omega({\log}n)$ space is reduced to $\Omega({\log}n{\log}{\log}n)$. The introduction of one global current vertex and always returning information about this vertex, mimics the implementation and correctness of Chong-Lam algorithm with minor modifications to the hooking scheme. The current vertex is an implicit argument to all functions describing a {\em{configuration}}.

To generalize this algorithm to \SGSLOGCFL, we make the following modifications : (i) add two hooking procedures (one for growing $s$-components and one for growing $g$-components) (ii) the new edges found after every call {\em{Contract}} are added in the matrices ${\Upsilon}^*$ and $\mathcal{E}^*$ and the new degrees of the vertices are recomputed and (iii) the exploration walks and the bit by bit comparison are done on the hooking trees generated by the $s$-components and $g$-components.

\begin{theorem}\label{thm:sgslogcfl-space-lognloglogn}
Let $\mathcal{G} = {\langle}{\Upsilon},\mathcal{E}{\rangle}$ be an instance of \SGUSTREAL. $\mathcal{G}^* = {\langle}{\Upsilon}^*,\mathcal{E}^*{\rangle}$ can be computed deterministically in $O({\log}{n}{\log}{\log}{n})$ space i.e., \SGSLOGCFL\ $\in$ $DSPACE({\log}{n}{\log}{\log}{n})$.
\end{theorem}

\begin{corollary}\label{cor:bstconn-space-lognloglogn}
\BSTCONN\ $\in$ $DSPACE({\log}{n}{\log}{\log}{n})$.
\end{corollary}

\section{Open Problems}

In a recent work \cite{kintali-complement}, we proved that \BSTCONN, \SGSLOGCFL\ and \PBSTCONN\ are all closed under complementation. Several interesting research directions arise from our work :

\begin{itemize}

\item{{\bf{Balanced Connectivity}}: \BSTCONN\ and \PBSTCONN\ are natural graph connectivity problems that lie between \logspace\ and \NL. Studying their space complexity is an interesting research direction towards improving the space complexity of \STCONN. In particular, it would be interesting to improve \thref{thm:sgslogcfl-space-lognloglogn}. Is \SGSLOGCFL\ $\in$ \logspace\ ? Less ambitiously, is \SGSLOGCFL\ $\in$ \SCtwo\ ?}

\item{An alternate proof of \thref{thm:sgslogcfl-space-lognloglogn} using the techniques of \cite{zigzag-journal, SL=L} or \cite{derand-squaring} seems to be a challenging task.}

\item{\SLOGCFL\ vs \LOGDCFL: In the logspace setting we have \logspace\ = \SL\ $\subseteq$ \NL. In the \LOGCFL\ setting, we have \LOGDCFL\ $\subseteq$ \SLOGCFL\ = \LOGCFL\ (see \thref{thm:all-lange-theorem}). By definition, we have \NL\ $\subseteq$ \LOGCFL. It is known that \LOGDCFL\ $\subseteq$ \SCtwo\ \cite{Cook-logDCFL}. This motivates the study of the relationship between \LOGDCFL\ and \SLOGCFL. It would be interesting to generalize the techniques of \cite{zigzag-journal, SL=L} to prove \LOGDCFL\ = \SLOGCFL. This would imply \NL\ $\subseteq$ \SCtwo, i.e., \STCONN\ can be solved by a deterministic algorithm in polynomial time and $O({\log}^2{n})$ space.}

\item{\SLOGCFL\ vs \RLOGCFL: We have \LOGDCFL\ $\subseteq$ \SLOGCFL\ = \LOGCFL\ and \LOGDCFL\ $\subseteq$ \allowbreak \RLOGCFL\ $\subseteq$ \LOGCFL\ implying \RLOGCFL\ $\subseteq$ \SLOGCFL. In the logspace setting, prior to Reingold's work, Aleliunas et. al. \cite{AKLLR} proved that \SL\ $\subseteq$ \RL, using random walks. It would be interesting to generalize their techniques to prove \SLOGCFL\ $\subseteq$ \RLOGCFL. Since \BPLOGCFL\ $\subseteq$ \SCtwo\ \cite{venkat-auxpda}, a proof of \SLOGCFL\ $\subseteq$ \RLOGCFL\ would imply \NL\ $\subseteq$ \SCtwo.}

\item{Is there a circuit characterization of \SGSLOGCFL\ ? What is the relationship between (i) \SGSLOGCFL\ and \NL\ ? (ii) \SGSLOGCFL\ and \LOGDCFL\ ? (iii) \SGSLOGCFL\ and ${\bf{DET}}$\footnote{${\bf{DET}}$ is the class of problems \NCone\ Turing reducible to the determinant \cite{Cook85-DET}.} ?}

\item{Allender and Lange \cite{allender-lange} proved that \SLOGCFL\ = \LOGCFL. Is \oneSLOGCFL\ = \oneLOGCFL\ ? i.e., is \PBSTCONN\ \NL-complete ?}


\end{itemize}

\vspace{0.2in}

\noindent {\large{\bf{Acknowledgements}}} : This project is partially funded by the NSF grant CCF-0902717. I gratefully acknowledge helpful discussions with Eric Allender, Klaus-J{\"o}rn Lange, Nutan Limaye, Richard J. Lipton, H. Venkateswaran and Dieter van Melkebeek.

\bibliographystyle{alpha}
\bibliography{../bib-kintali}

\newpage
\vspace{0.15in}
\noindent {\bf{\LARGE{Appendix}}}

\appendix

\section{Symmetric AuxPDAs}\label{sec:symmauxpda}

An {\bf{auxiliary pushdown automaton}} (AuxPDA) is a multi-tape Turing machine with a two-way read-only input tape, a pushdown tape, and one or more work tapes. The pushdown alphabet has a distinguished symbol (say $\$$) which is initially {\em{pushed}} on the pushdown tape. The machine is designed so that the pushdown head never shifts left of $\$$ or changes $\$$. Further, the pushdown head can never shift left when scanning any tape symbol unless it first erases (i.e., pops) that symbol, and it can never shift right from a square unless it first prints (i.e., pushes) a nonblank symbol on that square. Space on an AuxPDA is the space used on the work tapes without counting the space on the pushdown tape. Formally, an AuxPDA is an 8-tuple $M = (Q, \Sigma, \Sigma_0, \Sigma_{\alpha}, l, \Delta, s, F)$, where $Q$ is a finite set of states, $\Sigma$ is a finite tape alphabet, $\Sigma_0 \subseteq \Sigma$ is the input alphabet, $\Sigma_{\alpha} \subseteq \Sigma$ is the pushdown alphabet, $l$ is the number of tapes, $s \in Q$ is the initial state, $F \subseteq Q$ is the set of final states and $\Delta$ is a finite set of transitions.

We first define the {\em{transition}} of an AuxPDA that enable the AuxPDA to ``peek" one square right or left on the input and work tapes and one square below the top symbol of the pushdown tape while changing its configuration. A transition is of the form $(p, \mathcal{S}, t_1, \dots, t_l, q)$, where $p$ and $q$ are states, $\mathcal{S}$ is a stack triple, $l$ is the number of tapes, and $t_1, \dots, t_l$ are tape triples.  A stack triple is either of the form (i) $({\alpha_a}{\alpha_b}, P, {\alpha_c}{\alpha_d})$, where ${\alpha_a},{\alpha_b},{\alpha_c},{\alpha_d} \in \Sigma_{\alpha}$ and $P$ is +1 or -1 ; or is of the form (ii) $({\alpha_a},0,{\alpha_b})$, where ${\alpha_a},{\alpha_b} \in \Sigma_{\alpha}$. A tape triple is either of the form (i) $(ab, D, cd)$, where $a, b, c, d \in \Sigma$ and $D$ is +1 or -1; or is of the form (ii) $(a,0,b)$, where $a, b \in \Sigma$.

A transition of the form $(p, \mathcal{S}, t_1, \dots, t_l, q)$ signifies that $M$ moves from state $p$ to state $q$ according to the stack and tape triples. The tape triple $t_i = (ab, +l, cd)$ signifies that when $M$ is scanning symbol $a$ on tape $t_i$, and with the square just to the right of the scanned square containing symbol $b$, $M$ may rewrite these two squares to contain symbols $c$ and $d$, respectively, move its tape head one square to the right. Similarly, a transition $(ab, -1, cd)$ signifies a potential left movement of the tape head, except that now the scanned symbol must be $b$ and the one to its left $a$ and these are rewritten as $d$ and $c$, respectively. The tape triple $(a,0,b)$ signifies that $M$ replaces the symbol $a$ with $b$ without moving its head position. The stack triple is defined analogously with $P=+1$ (resp. $P=-1$) corresponding to a push (resp. pop) operation on the pushdown tape.

The {\bf{surface configuration}} (introduced by Cook \cite{cook-auxpda}) of an AuxPDA on an input $w$ consists of the state, contents and head positions of the work tapes, the head position of the input tape and the topmost symbol of the stack. Note that for a space $S(n)$-bounded AuxPDA, its surface configurations take only $O(S(n))$ space. In the rest of this section, we will refer to surface configurations as configurations. Let $\mathcal{C}(M)$ denote the set of all configurations of $M$. For an input $w$, and $C_1, C_2 \in \mathcal{C}(M)$ we write $C_1\ {\vdash}_M\ C_2$ to denote that $C_1$ ``yields" $C_2$. A {\em{computation}} by $M$ is a sequence $C_0\ {{\vdash}_M}\ C_1{{\vdash}_M}\ {\dots}\ {{\vdash}_M}\ C_n$, where $n \geq 0$ and $C_0,\dots,C_n \in \mathcal{C}(M)$. The reflexive, transitive closure of ${\vdash}_M$ is denoted by $\displaystyle{\vdash}^*_M$ and the transitive closure is denoted by $\displaystyle{\vdash}^+_M$. An AuxPDA $M$ is nondeterministic (resp. deterministic) if ${\vdash}_M$ is multi-valued (resp. single-valued).

Since the tape triples and stack triples of $M$ enable it to peek into only a {\em{constant}} number of symbols, $M$ can be simulated by a standard AuxPDA extending the notion of {\em{big-headed}} Turing machines \cite{bigheaded-book}. The ``peeking" version of $M$ enables us to define symmetric computation. Each transition $\delta = (p, \mathcal{S}, t_1, \dots, t_l, q)$ has an inverse $\delta^{-1} = (q, \mathcal{S}^{-1}, t_1^{-1}, \dots, t_l^{-1}, p)$ where if $\mathcal{S} = (\alpha,P,\beta)$ then $\mathcal{S}^{-1} = (\beta,-P,\alpha)$ and for $i=1,\dots,k$ if $t_i = (a,D,b)$ then ${t_i}^{-1} = (b,-D,a)$.

The inverse of an AuxPDA $M = (Q, \Sigma, \Sigma_0, \Sigma_{\alpha}, l, \Delta, s, F)$ is $M^{-1} = (Q, \Sigma, \Sigma_0, \Sigma_{\alpha}, l, \Delta^{-1}, s, F)$, where $\Delta^{-1} = \{\delta^{-1} : \delta \in \Delta \}$. An AuxPDA is {\em{symmetric}} if it is its own inverse i.e., if $\delta^{-1} \in \Delta$ whenever $\delta \in \Delta$. The symmetric closure of an AuxPDA $M = (Q, \Sigma, \Sigma_0, \Sigma_{\alpha}, l, \Delta, s, F)$ is $\overline{M} = (Q, \Sigma, \Sigma_0, \Sigma_{\alpha}, l, {\Delta}{\cup}{\Delta}^{-1}, s, F)$. Note that the symmetric closure of an AuxPDA is symmetric and  a symmetric AuxPDA is its own symmetric closure. We now define the complexity class \SLOGCFL.

\begin{framed}
\noindent \SLOGCFL\ is the class of languages accepted by log space bounded and polynomial time bounded symmetric AuxPDA.
\end{framed}

Let $\#$ be a new special symbol in the tape alphabet that does not belong to input alphabet. For an AuxPDA $M$, $M^{\#}$ is its normal form such that (1) $M$ and $M^{\#}$ accept the same language in the same space bound, and have the same number of tapes; (2) $M^{\#}$ has no transitions into its initial state or out of any final state; (3) for any configurations $C_1,C_2 \in \mathcal{C}(M^{\#})$ if $C_1 {\vdash}_{M^{\#}} C_2$ then $|C_1|\leq|C_2|$, where $|C|$ represents the space of $C$. $M^{\#}$ is constructed from $M$ by adding a new initial state and transitions from it to the old initial state; eliminating any transitions out of final states; and introducing a new {\em{pseudoblank}} symbol which $M^{\#}$ writes instead of writing (or rewriting) a blank on a worktape, and which $M^{\#}$ treats as indistinguishable from a blank when seen on a worktape. $\overline{M^{\#}}$ is the symmetric closure of $M^{\#}$.

The following lemma is proved by Lewis and Papadimitriou \cite{SymmLogspace} in the context of symmetric Turing machines. By our definition of symmetric AuxPDA's, its proof follows by treating the ``configurations" of a symmetric Turing machine as the ``surface configurations" of a symmetric AuxPDA and augmenting the transitions with stack triples. We skip its proof since it is essentially the proof of \cite{SymmLogspace}.

\begin{lemma}\label{lem:SL-sub-lemma}
Let $M = (Q, \Sigma, \Sigma_0, \Sigma_{\alpha}, l, \Delta, s, F)$ be any AuxPDA, and let $\mathcal{A} \subseteq \mathcal{C}(M)$. Suppose that \\
\indent (a) for any $A_1, A_2 \in \mathcal{A}$, if $A_1\ \displaystyle{\vdash}^{+\mathcal{A}}_M\ A_2$ then $A_2\ \displaystyle{\vdash}^{+\mathcal{A}}_M\ A_1$ \\
\indent (b) for any $A \in \mathcal{A} \cup \mathcal{I}(M)$, and $B \notin \mathcal{A}$ and any $C_1,C_2,C_3$, if $A\ \displaystyle{\vdash}^{*\mathcal{A}}_M\ C_1\ \displaystyle{\dashv}^{*\mathcal{A}}_M\ C_2\ {\dashv}_M\ B\ {\vdash}_M\ C_3$, then $C_2 = C_3$ \\
\indent (c) for any $A_1 \in \mathcal{A} \cup \mathcal{I}(M)$, any $A_2 \in \mathcal{A}$, and any $B$, if $A_1\ \displaystyle{\vdash}^{*\mathcal{A}}_M\ B\ \displaystyle{\dashv}^{*\mathcal{A}}_M\ A_2$ then $A_1 = A_2$. \\
Then $\overline{M^{\#}}$ accepts the same language as $M$ in the same space as $M$.
\end{lemma}

\noindent \thref{thm:gen-SL-inclusion} \LOGDCFL\ $\subseteq$ \SLOGCFL\ $\subseteq$ \LOGCFL.

\begin{proof}
Let $M$ be a deterministic logspace bounded AuxPDA accepting a language $L \in\ $\LOGDCFL. Then $M$ satisfies the hypothesis of \lemref{lem:SL-sub-lemma}, with $\mathcal{A} = \emptyset$. $M$ satisfies the hypothesis (a) and (c) trivially. Since $M$ is deterministic it satisfies the hypothesis (b). Hence $\overline{M^{\#}}$ accepts $L$. Hence, \LOGDCFL\ $\subseteq$ \SLOGCFL. The second inclusion is trivial, since nondeterminism is more general than symmetry.
\end{proof}

Now that we have the definition and properties of \SLOGCFL, the proofs of the following theorem and its corollary are similar to those of \thref{thm:streal} and \corref{cor:strealnoepsilon}. It is routine to check that the AuxPDA thus constructed, satisfies the properties of \lemref{lem:SL-sub-lemma}. \\

\noindent \thref{thm:ustreal} \USTREAL\ is \SLOGCFL-complete. \\

\noindent \corref{cor:ustrealnoepsilon} \USTREAL\ with no $\epsilon$ edges is \SLOGCFL-complete.

\section{Proofs}

\noindent \thref{thm:streal} \STREAL\ is \LOGCFL-complete.
\begin{proof}
We first show that \STREAL\ is in \LOGCFL. Let ${\langle}\mathcal{G}(V,E),s,t{\rangle}$ be an instance of \STREAL. An AuxPDA (say $\mathcal{M}$) deciding \STREAL\ operates by starting at node $s$ and nondeterministically guessing the nodes of a directed path from $s$ to $t$. $\mathcal{M}$ records the position of the current node at each step on the work tape. If the current node is $u$, $M$ nondeterministically selects the next node $v$ such that $(u,v)$ is a directed edge in $H$. Let the labels of $u$ and $v$ be $\alpha_u$ and $\alpha_v$ respectively. If $(u,v)$ is labeled $push$, $\mathcal{M}$ pushes $\alpha_v$ onto its pushdown tape. If $(u,v)$ is labeled $pop$, it pops $\alpha_u$ from its pushdown tape and verifies that the new symbol on the stack is $\alpha_v$. If not, it terminates and $rejects$. If $(u,v)$ is labeled $\epsilon$ then $\mathcal{M}$ checks if $\alpha_u$ and $\alpha_v$ are equal. If not, it terminates and $rejects$. $M$ repeats this action until it reaches node $t$ with an empty pushdown tape and $accepts$, or until it has gone on for $N$ steps and $rejects$, where $|V|=N$ is the number of nodes in $\mathcal{G}$. Hence \STREAL\ is in \LOGCFL.

We now show a log space reduction from any language $\mathcal{L}$ in \LOGCFL\ to \STREAL. Let $\mathcal{M}$ be an AuxPDA deciding $\mathcal{L}$ in log space. Given an input $w$, we construct a directed graph $\mathcal{H}$ along with the vertex and labels and two special vertices $s$ and $t$ such that $\mathcal{H}$ has a {\em{realizable}} path from $s$ to $t$ if and only if $\mathcal{M}$ accepts $w$.

The nodes of $\mathcal{H}$ are the configurations of $\mathcal{M}$ on $w$. For configuration $A$ and $B$ of $\mathcal{M}$ on $w$, the pair $(A,B)$ is an edge of $\mathcal{H}$ if $B$ is one of the possible next configurations of $M$ starting at $A$. We say that $A$ {\em{yields}} $B$. The vertices of $\mathcal{H}$ are labeled with the topmost symbol of the stack in the corresponding configuration of $\mathcal{M}$. The edge $(A,B)$ is labeled $push$ (resp. $pop$) if $\mathcal{M}$ performs a $push$ (resp. $pop$) operation to reach from $A$ to $B$. If $\mathcal{M}$ reaches from $A$ to $B$ without a push or pop then the edge $(A,B)$ is labeled $\epsilon$. Node $s$ is the start configuration of $\mathcal{M}$ on $w$. We may assume that $\mathcal{M}$ has a unique accepting configuration, and we designate this configuration to be node $t$. This mapping reduces $\mathcal{L}$ to \STREAL\ because, whenever $M$ accepts $w$, some branch of its computation accepts, which corresponds to a realizable path from $s$ to $t$ in $\mathcal{H}$. Conversely, if some realizable path exists from $s$ to $t$ in $\mathcal{H}$, some computation branch accepts when $\mathcal{M}$ runs on input $w$. The reduction can be performed by a log space transducer which, on input $w$, outputs a description of $\mathcal{H}$ along with the vertex and edge labels.
\end{proof}

\vspace{0.2in}

\noindent \corref{cor:strealnoepsilon} \STREAL\ with no $\epsilon$ edges is \LOGCFL-complete.
\begin{proof}
We replace each directed edge $(u,v)$ labeled with $\epsilon$ with two directed edges $(u,w)$ and $(w,v)$, where $w$ is a new node. The label of $(u,w)$ (resp. $(w,v)$) is set to $push$ (resp. $pop$). Repeat this for every edge, adding a new node every time. We introduce a new label $\alpha_{k+1}$ and label all the new nodes with $\alpha_{k+1}$. It is easy to see that a path from $s$ to $t$ is realizable in the original graph if and only if it is realizable in the new graph. Hence \STREAL\ reduces to \STREAL\ with no $\epsilon$ edges.

Equivalently, we may assume that an AuxPDA always pushes or pops a symbol at every step. If an AuxPDA doesn't push or a pop at every step then we introduce an extra alphabet in its stack alphabet which is pushed onto the stack when nothing is done to the stack. This alphabet is first popped before performing a valid stack move.
\end{proof}

\vspace{0.2in}

\noindent \thref{thm:nl-1logcfl} \NL\ = \oneLOGCFL.
\begin{proof}
\oneLOGCFL\ $\subseteq$ \NL: An \NL-machine (say $\mathcal{M}$) non-deterministically guesses an $s$-$t$ path (say $P$). $\mathcal{M}$ traverses the edges along $P$ and maintains a counter $C$. $\mathcal{M}$ increments (resp. decrements) $C$ if the current edge is labeled $push$ (resp. $pop$). If $C$ was ever negative then $\mathcal{M}$ rejects. $\mathcal{M}$ accepts iff $C=0$ when it reaches $t$. \\

\noindent \NL\ $\subseteq$ \oneLOGCFL : We replace each directed edge (say $(u,v)$) of \STCONN\ by two directed edges $(u,w)$ and $(w,v)$ and label them $push$ and $pop$ respectively. We add a new vertex $w$ for each edge $(u,v)$. There is an $s$-$t$ path in the original graph iff there is a realizable path (according to the definition from \secref{one-logcfl}) in the modified graph.
\end{proof}

\vspace{0.2in}

\noindent \thref{thm:1sgslogcfl} \BSTCONN\ is \oneSGSLOGCFL-complete.
\begin{proof}
\BSTCONN\ $\in$ \oneSGSLOGCFL: Let $\mathcal{G}(V,E)$ be an instance of \BSTCONN. Let $\mathcal{G'}(V,E')$ be the underlying undirected graph of $\mathcal{G}$. If $(u,v) \in E$ and $(v,u) \in E$ then we label the edges $(u,v)$ and $(v,u)$ of $\mathcal{G'}$ with $\epsilon$. If $(u,v) \in E$ and $(v,u) \notin E$ then we label the edge $(u,v)$ of $\mathcal{G'}$ with $push$ and label the edge $(v,u)$ of $\mathcal{G'}$ with $pop$. Note that the edge labels of $\mathcal{G'}$ are symmetric. There is a balanced $s$-$t$ path in $\mathcal{G}$ iff there is a realizable $s$-$t$ path (according to the definition from \secref{one-sgslogcfl}) in $\mathcal{G'}$.

\noindent \BSTCONN\ is \oneSGSLOGCFL-hard: An instance of \oneSGSLOGCFL\ is an undirected graph (say $G$) with edges labeled from $\{push,pop,\epsilon\}$. These edge labels are symmetric as defined in \secref{sec:ustconn}. We construct a directed graph $H$ on the same vertex set. If the edge $(u,v)$ of $G$ is labeled $\epsilon$ we add the edges $(u,v)$ and $(v,u)$ in $H$. If the edge $(u,v)$ is labeled $push$ (by symmetry the edge $(v,u)$ is labeled $pop$) we add a directed edge $u,v$ in $H$. There is a realizable $s$-$t$ path in $G$ iff there is a balanced $s$-$t$ path in $H$. \\
\end{proof}

\vspace{0.2in}

\noindent \thref{thm:1slogcfl} \PBSTCONN\ is \oneSLOGCFL-complete.
\begin{proof}
Similar to the proof of \thref{thm:1sgslogcfl}.
\end{proof}

\vspace{0.2in}

\noindent \thref{thm:ckt-square} Let $\mathcal{G}$ be an instance of \STREAL. $\mathcal{G}^* = {\langle}{\Upsilon}^*,\mathcal{E}^*{\rangle}$ can be computed using $O({\log}n)$ repeated applications of ${\bf{Square}}(\mathcal{G})$.
\begin{proof}
We first state the relevant definitions and lemmas from \cite{NiedermeierR95}. A {\em{path description}} is a triple $(A, B, i)$ consisting of two surface configurations $A$ and $B$ and an even natural number $i$. A description is {\em{realizable}} if $A$ and $B$ are realizable. By Corollary \ref{cor:strealnoepsilon} we may assume that there are no $\epsilon$ edges in an instance of \STREAL, and hence $i$ can only be an even number. In particular, $(A, B, i)$ represents several paths of length $i$ between $A$ and $B$.

The relation $\vdash$ shows how to split computation paths recursively into shorter and shorter paths until we end up with trivial paths. Let $x = (A,B,i)$, $y = (C,D,j)$, and $z = (E,B,k)$ be path descriptions. Then we write $y, z\ {\vdash}\ x$ and $z,y\ {\vdash}\ x$ if and only if \\
\indent (1) the level of the pushdown is equal for $A$, $E$ and $B$; \\
\indent (2) there exists a computation from $A$ to $C$ in one step, pushing a symbol $\alpha$ onto the pushdown tape during this step; \\
\indent (3) there exists a computation from $D$ to $E$ in one step, popping $\alpha$ from the pushdown tape; and \\
\indent (4) $j + k = i - 2$.

Note that identical pushdown heights of $A$, $E$ and $B$ imply that $C$ and $D$ have same pushdown height. Also, $j$ and $k$ are always even. In this way we can reduce the checking of realizability of $x$ to the checking of the realizability of smaller paths $y$ and $z$. We now state two crucial lemmas from \cite{NiedermeierR95} that gives a ``balanced" partition of realizable computation. The proofs of these lemmas are based on a {\em{recursive descent}} using the properties of the decomposition relation ${\vdash}$.

\begin{lemma}\label{lem:NR-lemma-standard} (Niedermeier and Rossmanith \cite{NiedermeierR95})
Let $(A,B,i)$ denote a realizable path description for a fixed computation path of length $i \geq 2$ between $A$ and $B$. Then there exist uniquely determined subpaths $(C, D, i_1)$, $(E, F, i_2)$ and $(G, D, i_3)$ of $(A, B, i)$ such that $(E, F, i_2), (G, D, i_3)\ {\vdash}\ (C, D, i_1)$ and $i_2, i_3 \leq i/2 < i_1$.
\end{lemma}

\lemref{lem:NR-lemma-standard} splits a fixed computation path into three paths. The first two paths are the subpaths $(E,F,i_2)$ and $(G,D,i_3)$ and the third one is the path $(A,B,i)$ {\em{with gap}} $(C,D,i_1)$. This means that the verification of the realizability of $(A,B,i)$ can be reduced to showing that $(E, F, i_2)$, $(G, D, i_3)$ and the pair-with-gap $(A,(C,D,i_1),B,i)$ are realizable. 

A description for a {\em{path with gap}} $(A,(C,D,j),B,i)$ consists of four surface configurations $A, B, C, D$ and two even numbers $i$ and $j$ with $j \leq i$. A path with gap $(A,(C,D,j),B,i)$ is called {\em{realizable}} iff $(A{\leadsto}(C,D){\leadsto}B)$ and there exists a computation path from $A$ to $C$ and one from $D$ to $B$ with total number of steps $j-i$. Now we generalize the decomposition relation ${\vdash}$ to computation paths with gap. Let $x = (A,(C,D,j),B,i)$ and, first, let $y = (E,(C,D,j),F,k)$ and $z = (G,B,l)$ or, second, let $y = (E,F,k)$, $z = (G,(C,D,j),B,l)$. Then we write $y, z\ {\vdash}\ x$ and $z,y\ {\vdash}\ x$ if and only if \\
\indent (1) the level of the pushdown is equal for $A, G$ and $B$; \\
\indent (2) there exists one step from $A$ to $E$ pushing a symbol $\alpha$ onto the pushdown tape; \\
\indent (3) there is one step from $F$ to $G$ popping $\alpha$ from the pushdown tape; and \\
\indent (4) $k + l = i - 2$. \\

The following lemma is the analogue to \lemref{lem:NR-lemma-standard} for a fixed computation path with gap.

\begin{lemma}\label{lem:NR-lemma-gap} (Niedermeier and Rossmanith \cite{NiedermeierR95})
Let $(A, (C, D, j), B, i)$, $i-j \geq 2$ denote a realizable path with gap. Then there exist uniquely determined paths $y = (E, (C, D, j), F, i_1)$ and either \\
\indent (1) $z_1 = (G, (C, D, j), H, i_2)$ and $z_2 = (I, F, i_3)$, such that $z_1, z_2\ {\vdash}\ y$ and $i_2-j \leq (i-j)/2 < i_1-j$ or \\
\indent (2) $z_1 = (G, H, i_2)$ and $z_2 = (I, (C, D, j), F, i_3)$, such that $z_1,z_2\ {\vdash}\ y$ and $i_3-j \leq (i-j)/2 < i_1-j$.
\end{lemma}

\lemref{lem:NR-lemma-gap} is used to decompose {\em{paths with gaps}} in a balanced way. To check the realizability of $(A,(C,D,j),B,i)$ we examine the realizability of $(A,(E,F,i_1),B,i)$, $z_1$ and $z_2$. Both possible subpaths with gap have length less than or equal to half of the lenght of the whole path with gap $(A,(C,D,j),B,i)$. The arising subpath without gap may have a maximum length of $i-j-2$ and will be split in a balanced way using \lemref{lem:NR-lemma-standard}.

\noindent \line(1,0){450} \\
\noindent ${\bf{Square}}({\langle}{\Upsilon},\mathcal{E}{\rangle})$ \\

\indent for all $a,b \in V$ update $\mathcal{E}$ as follows :
\begin{eqnarray*}
\mathcal{E}[a,b] & = & \displaystyle\sum_{c,e,f,g,d}{\Upsilon}[a,(c,d),b]{\cdot}{\Upsilon}[c,(e,f),g]{\cdot}\mathcal{E}[e,f]{\cdot}\mathcal{E}[g,d] \\
\end{eqnarray*}
\indent for all $a,b,c,d \in V$ update ${\Upsilon}$ as follows :
\begin{eqnarray*}
{\Upsilon}[a,(c,d),b] & = & \displaystyle\sum_{c',e',f',g',d'}{\Upsilon}[a,(c'd'),b]{\cdot}{\Upsilon}[c',(e',f'),g']{\cdot}{\Upsilon}[e',(c,d),f']{\cdot}\mathcal{E}[g',d'] \\
& + & \displaystyle\sum_{c',e',f',g',d'}{\Upsilon}[a,(c'd'),b]{\cdot}{\Upsilon}[c',(e',f'),g']{\cdot}\mathcal{E}[e',f']{\cdot}{\Upsilon}[g',(c,d),d'] \\
\end{eqnarray*}
\indent {\bf{return}} ${\langle}{\Upsilon},\mathcal{E}{\rangle}$ \\
\noindent \line(1,0){450} \\

\noindent \line(1,0){300} \\
\noindent ${\bf{Square}}({\langle}{\Upsilon},\mathcal{E}{\rangle})$ \\
\indent $\mathcal{E} = (\Upsilon \otimes_2((\Upsilon \otimes_2 \mathcal{E})\otimes_1 \mathcal{E}))$ \\
\indent $\Upsilon = (\Upsilon \otimes_5((\Upsilon \otimes_5 \Upsilon)\otimes_3 \mathcal{E})) 
                         + (\Upsilon \otimes_5((\Upsilon \otimes_2 \mathcal{E})\otimes_4 \Upsilon)) $\\
\indent {\bf{return}} ${\langle}{\Upsilon},\mathcal{E}{\rangle}$ \\
\noindent \line(1,0){300} \\

Our {\bf{Square}} algorithm is based on \lemref{lem:NR-lemma-standard} and \lemref{lem:NR-lemma-gap}. Since the summation is taken over all possible intermediate surface configurations, the matrices $\mathcal{E}$ and $\Upsilon$ are populated in a {\em{bottom-up}} manner. Based on the above lemmas, Niedermeier and Rossmanith \cite{NiedermeierR95} constructed an \SACone\ circuit simulating the corresponding AuxPDA. The circuit consists of gates denoted by $\langle A, B, i\rangle$ and $\langle A, (C, D, j), B, i\rangle$ that compute the realizability of the corresponding path descriptions. Our ${\bf{Square}}$ algorithm is inspired by their approach. Translating the sum symbols into (unbounded) OR-gates and multiplication symbols into (bounded) AND-gates we get the corresponding \SACone\ circuit for a given vertices $s$ and $t$ of the graph $\mathcal{G}$. Each {\bf{Square}} operation on the graph $\mathcal{G} = {\langle}{\Upsilon},\mathcal{E}{\rangle}$ reduces the depth of the corresponding circuit by $O(1)$. Since our squaring operation is used to update all the entries of $\mathcal{E}$ and $\Upsilon$, after $O({\log}n)$ repeated squaring operations, we can decide the $s$-$t$ realizability for any two given vertices $s$ and $t$. Similar argument holds for paths with gap.

Hence, we can compute the transitive closure $\mathcal{G}^* = {\langle}{\Upsilon}^*,\mathcal{E}^*{\rangle}$ using $O({\log}n)$ repeated squaring operations on $\mathcal{G} = {\langle}{\Upsilon},\mathcal{E}{\rangle}$.
\end{proof}

\vspace{0.2in}

\noindent \thref{thm:simple-square} Let $\mathcal{G}$ be an instance of \STREAL. $\mathcal{G}^* = {\langle}{\Upsilon}^*,\mathcal{E}^*{\rangle}$ can be computed using $O({\log}n)$ repeated applications of ${\bf{SimpleSquare}}(\mathcal{G})$.
\begin{proof}
For realizable paths (both standard and gap paths) of length at most four, it is easy to verify that an application of {\bf{SimpleSquare}} reduces the path length by a factor of at least $\frac{3}{4}$. For paths of length greater than four, we divide the path into three smaller paths using \lemref{lem:NR-lemma-standard} for standard paths and \lemref{lem:NR-lemma-gap} for path with gaps and use induction. This implies that one applcation of {\bf{SimpleSquare}} reduces the path length by a constant factor. Hence $O({\log}n)$ repeated applications of ${\bf{SimpleSquare}}(\mathcal{G})$ suffice to compute the transitive closure $\mathcal{G}^*$.
\end{proof}

\vspace{0.2in}

\noindent \thref{thm:parallel-log2n}
The algorithm {\bf{Connect}} finds $\mathcal{G}^* = {\langle}{\Upsilon}^*,\mathcal{E}^*{\rangle}$ in parallel time $O({\log}^2n)$ using $n^4$ processors in the CREW PRAM model.
\begin{proof}
The following observations state that the hooking process creates pseudotrees on vertices from $V$ and $V^2$. \\

\noindent {\bf{Observation}} : Let $V_s \subseteq V$ denote an $s$-component of $\mathcal{G}$ such that $|V_s| \geq 2$ and define the function $C : V_s \rightarrow V_s$ by $C(i) = {{\bf{StandardHook}}(i)}$. The function $C$ defines a directed graph $G_s(C) = (V_s,F)$ where $F = \{(i,C(i))\ |\ i \in V_s\}$. Then $G_s(C)$ is a collection of pseudotrees with circumference one, and the smallest-numbered vertex in each pseudotree is in the cycle of the pseudotree. \\

\noindent {\bf{Observation}} : Let $V_g \subseteq V^2$ denote a $g$-component of $\mathcal{G}$ such that $|V_g| \geq 2$ and define the function $C : V_g \rightarrow V_g$ by $C(i,j) = {{\bf{GapHook}}(i,j)}$. The function $C$ defines a directed graph $G_g(C) = (V_g,F)$ where $F = \{((i,j),C(i,j))\ |\ (i,j) \in V_g\}$. Then $G_g(C)$ is a collection of pseudotrees with circumference one, and the lexicographically smallest vertex in each pseudotree is in the cycle of the pseudotree. \\

The hooking processes ({\bf{StandardHook}} and {\bf{GapHook}}) and the contraction step are implemented to mimic the functionality of {\bf{SymmetricSquare}}. Hence the correctness of the contraction step and the overall algorithm follows from \corref{cor:symmetric-square}.

\noindent {\bf{Time and Processor Bounds}} : The main loop of the {\bf{Connect}} program is executed $O(\log{n})$ times. Within the loop, the iteration at step 14 is executed $O(\log{n})$ times. Thus the algorithm requires $\Omega({\log}^2{n})$ time. Steps 3, 12, 18 require $O(1)$ time using $\Omega(n)$ processors. Steps 4, 13, 19 require $O(1)$ time using $\Omega(n^2)$ processors. Steps 14-17 require $O({\log}{n})$ time using $\Omega(n^2)$ processors. {\bf{StandardHook}} and {\bf{GapHook}} are essentially computing minimum of at most $O(n^2)$ integers (accessing both $\mathcal{E}$ and $\Upsilon$) and hence can be programmed to execute in $O({\log}{n})$ time using $O(n^2)$ processors. Hence the total running time is $O({\log}^2{n})$. The total number of processors used is $O(n^4)$.
\end{proof}





\section{More details of \BSTCONN}\label{sec:bstconn}

\begin{figure}[htp]
\begin{center}
\includegraphics[width=4in]{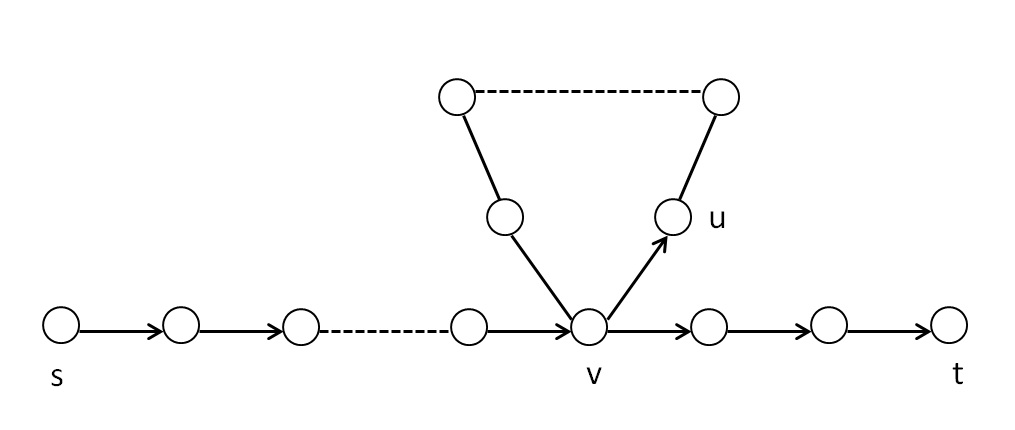}
\end{center}
\caption{Example.}
\label{fig:bstconn-n2}
\end{figure}

In all the algorithms presented in this paper we are only looking for balanced paths of length at most $n$. Our algorithms can easily be extended to find balanced paths of length $n^r$ where $r$ is explicitly specified as part of the input. The example in Figure \ref{fig:bstconn-n2} shows an instance of \BSTCONN\ where the {\em{only}} balanced path between $s$ and $t$ is of length $\Theta(n^2)$. The directed simple path from $s$ to $t$ is of length $n/2$. There is a cycle of length $n/2$ at the vertex $v$. All the edges (except $(v,u)$) on this cycle are undirected. The balanced path from $s$ to $t$ is obtained by traversing from $s$ to $v$, traversing the cycle clockwise for $n/2$ times and then traversing from $v$ to $t$. This path is not simple.

We now define two more connectivity problems. A path $P$ from $s$ to $t$ is called {\em{k-balanced}} if the number of forward edges along $P$ minus the number of backward edges along $P$ is equal to $k$. A path $P$ from $s$ to $t$ is called {\em{positive k-balanced}} if $P$ is {\em{positive}} and {\em{k-balanced}}.

\begin{framed}
\noindent \kBSTCONN\ : Given a directed graph $\mathcal{G}(V,E)$ and two distinguished nodes $s$ and $t$, decide if there is {\em{k-balanced}} path (of length at most $n$) between $s$ and $t$.
\end{framed}

\begin{framed}
\noindent \kPBSTCONN\ : Given a directed graph $\mathcal{G}(V,E)$ and two distinguished nodes $s$ and $t$, decide if there is {\em{positive k-balanced}} path (of length at most $n$) between $s$ and $t$.
\end{framed}

\kBSTCONN\ can be solved as follows : Add a new vertex $t'$ and a directed path (with all new vertices) of length $k$ from $t'$ to $t$. Find a balanced path from $s$ to $t'$ in this modified graph. \kPBSTCONN\ can be solved similarly. \\


\end{document}